\documentclass[twocolumn]{aastex63}

\submitjournal{ApJ. Accepted on December 22, 2020}

\shorttitle{Bulge and disk colors in the cluster core and outskirts}
\shortauthors{Barsanti et al.}
\graphicspath{{./}{figures/}}

\begin{document}

\title{The colors of bulges and disks in the core and outskirts of galaxy clusters}

\correspondingauthor{Stefania Barsanti}
\email{stefania.jess@gmail.com}

\author{S. Barsanti}
\affiliation{Department of Physics and Astronomy, Macquarie University, NSW 2109, Australia}
\affiliation{Astronomy, Astrophysics and Astrophotonics Research Centre, Macquarie University, Sydney, NSW 2109, Australia}

\author{M. S. Owers}
\affiliation{Department of Physics and Astronomy, Macquarie University, NSW 2109, Australia}
\affiliation{Astronomy, Astrophysics and Astrophotonics Research Centre, Macquarie University, Sydney, NSW 2109, Australia}

\author{R. M. McDermid}
\affiliation{Department of Physics and Astronomy, Macquarie University, NSW 2109, Australia}
\affiliation{Astronomy, Astrophysics and Astrophotonics Research Centre, Macquarie University, Sydney, NSW 2109, Australia}

\author{K. Bekki}
\affiliation{ICRAR, The University of Western Australia, 35 Stirling Highway, Crawley, WA 6009, Australia}


\author{J. J. Bryant}
\affiliation{ARC Centre of Excellence for All Sky Astrophysics in 3 Dimensions (ASTRO 3D), Australia}
\affiliation{Sydney Institute for Astronomy (SIfA), School of Physics, University of Sydney, NSW 2006, Australia}
\affiliation{Australian Astronomical Optics, AAO-USydney, School of Physics, University of Sydney, NSW 2006, Australia}


\author{S. M. Croom}
\affiliation{ARC Centre of Excellence for All Sky Astrophysics in 3 Dimensions (ASTRO 3D), Australia}
\affiliation{Sydney Institute for Astronomy (SIfA), School of Physics, University of Sydney, NSW 2006, Australia}



\author{S. Oh}
\affiliation{ARC Centre of Excellence for All Sky Astrophysics in 3 Dimensions (ASTRO 3D), Australia}
\affiliation{Research School of Astronomy and Astrophysics, Australian National University, Canberra, ACT 2611, Australia}

\author{A. S. G. Robotham}
\affiliation{ICRAR, The University of Western Australia, 35 Stirling Highway, Crawley, WA 6009, Australia}
\affiliation{ARC Centre of Excellence for All Sky Astrophysics in 3 Dimensions (ASTRO 3D), Australia}

\author{N. Scott}
\affiliation{ARC Centre of Excellence for All Sky Astrophysics in 3 Dimensions (ASTRO 3D), Australia}
\affiliation{Sydney Institute for Astronomy (SIfA), School of Physics, University of Sydney, NSW 2006, Australia}
\author{J. van de Sande}
\affiliation{ARC Centre of Excellence for All Sky Astrophysics in 3 Dimensions (ASTRO 3D), Australia}
\affiliation{Sydney Institute for Astronomy (SIfA), School of Physics, University of Sydney, NSW 2006, Australia}



\begin{abstract}

 The role of the environment on the formation of S0 galaxies is still not well understood, specifically in the outskirts of galaxy clusters. We study eight low-redshift clusters, analyzing galaxy members up to cluster-centric distances $\sim2.5\,R_{200}$. We perform 2D photometric bulge-disk decomposition in the $g$-, $r$- and $i$-bands from which we identify 469 double-component galaxies. We analyze separately the colors of the bulges and the disks and their dependence on the projected cluster-centric distance and on the local galaxy density. For our sample of cluster S0 galaxies, we find that bulges are redder than their surrounding disks, show a significant color-magnitude trend, and have colors that do not correlate with environment metrics. On the other hand, the disks associated with our cluster S0s become significantly bluer with increasing cluster-centric radius, but show no evidence for a color-magnitude relation. The disk color-radius relation is mainly driven by galaxies in the cluster core at $0\leq R/ R_{200}<0.5$. No significant difference is found for the disk colors of backsplash and infalling galaxies in the projected phase space. Beyond $R_{200}$, the disk colors do not change with the local galaxy density, indicating that the colors of double-component galaxies are not affected by pre-processing. A significant color-density relation is observed for single-component disk-dominated galaxies beyond $R_{200}$. We conclude that the formation of cluster S0 galaxies is primarily driven by cluster core processes acting on the disks, while evidence of pre-processing is found for single-component disk-dominated galaxies.

We publicly release the data from the bulge-disk decomposition.  
\end{abstract}



\section{Introduction}
Environment is observed to play a strong role in galaxy evolution. In galaxy groups and clusters, galaxy properties such as morphology, star formation rate and color are found to correlate with the position of the galaxy within the halo and with the local galaxy density \citep{Dressler1980,Dressler1997,Lewis2002,Balogh2004}. According to the morphology-density relation of \citet{Dressler1980}, the fraction of lenticular (S0) galaxies in clusters is higher with respect to the fraction of spiral galaxies that mainly populate low density environments. This result indicates that S0 galaxies are important probes for understanding how the environment drives galaxy evolution.


The structural and kinematical properties of S0 galaxies and spirals are observed to be similar, therefore a possible scenario for forming S0s is through fading spiral galaxies \citep{vandenBergh1976,Bedregal2006,Moran2007,Prochaska2011}. The star formation activity of spiral galaxies is quenched when they enter the cluster environment, implying a possible transformation of these galaxies into S0s (e.g. \citealt{Dressler1997,Fasano2000,Kormendy2012}). These results suggest that environmental mechanisms acting within the clusters can contribute to the formation of S0 galaxies. However, since the bulges of S0s are observed to be different from those in spirals, cluster processes and subsequent fading of star formation cannot be solely responsible for the formation of S0 galaxies \citep{Dressler1980}.

Several environmental mechanisms in clusters have been observed as responsible for star formation quenching in spiral galaxies. Interactions with the intra-cluster medium cause a removal of the cold gas in the disk via ram pressure stripping \citep{Gunn1972}, or of the hot halo reservoir via strangulation \citep{Larson1980,Balogh2000}. Harrassment, due to tidal interactions with either the cluster potential \citep{Merritt1984, Mamon1987} or with neighbouring galaxies \citep{Moore1996}, can drive gas towards the centre of a galaxy triggering new star formation there.

\citet{Poggianti2009} analyzed the changing fraction of S0s relative to spirals as a function of redshift. They found that the evolution is stronger in lower mass halos. This suggests that pre-processing acting in groups and in the field is also important for the formation of S0 galaxies, before spirals are accreted onto the clusters. Mergers between galaxies are environmental mechanisms that mainly act in lower galaxy density environments \citep{Mihos1994}. Minor mergers can induce the transformation of a galaxy into an S0, for example by concentrating gas in the central galaxy regions to form a bulge \citep{Arnold2011,ElicheMoral2012}. However, the relative importance of S0 transformation in groups is still unclear compared to the galaxy transformation in the cores of clusters, where the effects of ram-pressure stripping are observed to act on infalling spirals \citep{Owers2019}. 

Isolated S0 galaxies can be the product of mechanisms acting within the galaxy. In situ starvation, due for example to disk instabilities or fragmentation, might cause field spiral galaxies to stop the star formation activity, lose the spiral structures and transform into S0s \citep{ElicheMoral2013,Saha2018}. Internal mechanisms of secular evolution within spiral galaxies can lead to the formation of pseudo-bulges from the disk gas \citep{Kormendy2004}.
 
S0 galaxies can be approximated as double-component
galaxies characterized by a central bulge and an outer
disk. The two components are believed to form through different formation processes, and different environmental mechanisms can also influence their star formation activity \citep{Hudson2010,Clauwens2018}. To understand completely the formation of S0 galaxies, it is necessary to study separately the properties of the bulge and the disk. The identification of S0 galaxies and their components using 1D or 2D photometric data is extensively used in the literature, since bulges and disks exhibit different projected surface brightness distributions. The light is fitted using S\'ersic models \citep{Sersic1963}. Statistical tests are applied to discern bulge/disk-dominated single-component galaxies from double-component galaxies where two components are actually needed to reproduce the data. Numerous catalogues have been published for low redshift galaxies \citep{Allen2006, Simard2011,Lackner2012,Mendel2014,Kelvin2014,Meert2015,Lange2016}, and recently also for high redshift galaxies \citep{Dimauro2018}.

The separate study of the bulge/disk colors offers insights into which environmental mechanisms are primarily responsible for the formation of S0 galaxies \citep{Hudson2010,Simard2011,Lackner2012,Lackner2013,Head2014}. Even if colors alone are not sufficient to disentangle the stellar population properties, such as age and metallicity, of the two components, they represent fossil records of the star formation histories. Bulge and disk colors offer a useful tool to understand if only one component or both components are responsible for the trends of the global galaxy color. The combination of the photometric information with spectroscopic data allows to inform about the ages and metallicities of bulges and disks for S0 galaxies in clusters and in the field \citep{Johnston2014,FraserMcKelvie2018}.   

\citet{Hudson2010} performed 2D photometric bulge-disk decomposition for eight nearby clusters. They found that bulges are redder and have a steeper color-magnitude relation than disks. The bulge colors are observed not to depend significantly on environment, while the disks colors in the cluster centre are found to be redder than those in the outskirts. They concluded that the star formation of bulges appears mainly due to internal galaxy processes, while the disk is also impacted by environmental mechanisms. \citet{Head2014} studied S0 galaxies in the Coma cluster. They found results in agreement with those of \citet{Hudson2010}, favoring a formation scenario for S0 galaxies where environmental mechanisms that cause disk fading, such as ram pressure stripping or strangulation, are primarily responsible. The analysis of both \citet{Hudson2010} and \citet{Head2014} is limited to galaxies within $R_{200}$, where most of the galaxy population is virialized. Going beyond $R_{200}$ allows us to trace infalling galaxies that are in groups or relatively isolated. At these larger cluster-centric distances, we can better explore the infalling populations and, therefore, investigate whether pre-processing plays an important role in the formation of S0 galaxies.

\citet{Lackner2013} studied 12,500 SDSS nearby galaxies in groups and clusters. They analyzed the bulge and disk properties as a function of environment metrics, such as local galaxy density, halo richness and crossing time. Oppositely to the results of \citet{Hudson2010} and \citet{Head2014}, they found that disk colors drive global color-density trends only for galaxies in poor groups. For rich clusters the disk colors were measured not to change with local density. \citet{Lackner2013} suggested that quenching of star formation occurs in groups, while the morphological transformation occurs in clusters. 

To understand the apparently conflicting results between  \citet{Hudson2010,Head2014} and \citet{Lackner2013} we investigate a sample of galaxy clusters that have deep, complete spectroscopy for defining environment metrics, but that also probe the regions beyond $R_{200}$ where pre-processing occurs. In this work we aim to assess how the environment affects the formation of S0 galaxies, assessing the importance of pre-processing compared to mechanisms acting in the cluster core. We study eight low-redshift massive galaxy clusters APMCC0917, A119, A168, A2399, A3880, A4038, A85 and EDCC0442, which belong to the Sydney-AAO Multi-Object Integral field (SAMI) cluster sample \citep{Croom2012,Bryant2015,Owers2017}. We select cluster members up to $\sim2.5\,R_{200}$, exploring the edge of the clusters and including the infalling regions. We perform 2D photometric bulge-disk decomposition in the $g$-, $r$- and $i$-bands to identify double-component galaxies. The decomposition makes use of the image analysis package {\sc ProFound} and the photometric galaxy profile fitting package {\sc ProFit} \citep{Robotham2017,Robotham2018}. We explore separately the bulge and disk colors as fossil records of their star formation properties. We investigate the dependence of the bulge and disk colors on the projected cluster-centric distance and on local galaxy density.  

This paper is organized as follows. We present our cluster sample and photometric data in Section~\ref{sec:Data}. In Section~\ref{2D bulge-disk decomposition} we describe the 2D bulge-disk decomposition to select double-component cluster galaxies. In Section~\ref{sec:Results} we present our results, discussing the comparison between the colors of the bulges and the disks. Finally, we summarize and conclude in Section~\ref{Summary and conclusions}. Throughout this work, we assume $\Omega_{m}=0.3$, $\Omega_{\Lambda}=0.7$ and $H_{0}=70$ km $\rm s^{-1} Mpc^{-1}$ as cosmological parameters.

\section{Data and cluster sample}
\label{sec:Data}
We study eight low-redshift massive galaxy clusters with virial masses $\log{(M_{200}/M_{\odot})}=14.25-15.19$ at $0.029<z<0.058$, which belong to the SAMI cluster sample \citep{Owers2017}. The data in \citet{Owers2017} were used to select the SAMI targets, but the cluster sample contains also galaxies with no SAMI observations.   

The four clusters APMCC0917, EDCC0442, A3880 and A4038 are selected from the 2dFGRS catalogue \citep{DePropris2002} with the photometry provided by the VLT Survey Telescope's ATLAS (VST/ATLAS) survey \citep{Shanks2013,Shanks2015}. The photometry used for the A85, A168, A119 and A2399 clusters in the SDSS regions is taken from SDSS DR9 \citep{Ahn2012}. The imaging data are used in Section~\ref{2D bulge-disk decomposition} for galaxy characterization with the purpose of selecting only double-component galaxies. A85 has coverage from both VST/ATLAS and SDSS imaging, but we use the SDSS data since only the SDSS survey completely covers the cluster. We present a comparison between the obtained double-component SDSS and VST/ATLAS samples in Section~\ref{Internal tests for 2D bulge-disk decomposition}.

For the selection of cluster members we use the catalogue of \citet{Owers2017}. Cluster membership is conducted consistently for the eight galaxy clusters by \citet{Owers2017}. They used an iterative approach that uses the combination of cuts in peculiar velocity, and caustic measurements in the projected phase space that tract the escape velocity profile to define membership. The cluster sample has a magnitude limit $r_{petro}\leq19.4$ and a high spectroscopic completeness of 94\%.

From the catalogue of \citet{Owers2017} we select cluster members with $M_{*}\geq10^{9.5}\,M_{\odot}$ and $R<2.5\,R_{200}$. The selection in $R/R_{200}$ is motivated by the cluster coverage. The selection in $M_{*}$ is due to less reliable photometric results for low mass galaxies and to the spectroscopic completeness limits (see Figure 7 of \citealp{Owers2017}). The total number of cluster members with SDSS imaging is 1233, while those with VST/ATLAS imaging is 605. We exclude 31 galaxy images with missing data due to their position close to a saturated star, the image edge or missing pixels. Visually inspecting the $r$-band images, we also exclude 12 massive and bright elliptical galaxies, which are the brightest cluster galaxies or are massive central galaxies in substructures. We exclude these massive and bright ellipticals since their complex light profiles can contaminate the double-component galaxy selection performed in Section~\ref{2D bulge-disk decomposition}. The final sample contains 1204 SDSS and 591 VST/ATLAS galaxy images, for a total of 1795 galaxies up to 2.5 $R_{200}$.

\section{Galaxy characterization}
\label{2D bulge-disk decomposition}
Our primary aim is to understand the bulge and disk colors of S0 galaxies in clusters. To do so, we must first identify those galaxies that can be decomposed into double-component systems. To that end, we perform two-dimensional (2D) photometric bulge-disk decomposition in the $g$-, $r$- and $i$-bands. For source finding and image analysis we use the {\sc ProFound} package \citep{Robotham2018}. {\sc ProFound} offers the key inputs and initial conditions to {\sc ProFit}, a package for Bayesian 2D photometric galaxy profile modelling \citep{Robotham2017}. We perform our primary structural decomposition using the $r$-band images. This is because the $r$-band is less susceptible to substructure due to ongoing star formation than the $g$-band, and it also has a higher signal-to-noise ratio than the redder $i$- and $z$-band data. We fit single-component S\'ersic profile and multiple double-component bulge+disk profiles for each galaxy in the $r$-band. We use these results as inputs for the single/double-component fits in the $g$- and $i$-bands. 

For galaxy characterization we use the Bayes Factor (BF), which compares the probabilities of two models $M1$ and $M2$ in reproducing the data $D$:
\begin{equation}
\ln(BF)=LML(D|M1)-LML(D|M2)
\end{equation}
where the log marginal likelihoods (LML) are obtained by integrating over the parameter space of each model. The advantage of using the BF is that it avoids over fitting, punishing each model for its number of parameters. Different criteria can be
used to estimate the strength of evidence of $M1$ versus $M2$ \citep{Jeffreys61,Kass1995}.

In this Section we describe the details of the 2D bulge-disk decomposition in the \textit{r}-, $g$- and $i$-bands. We select double-component galaxies and we perform internal tests to check the reliability of the galaxy characterization.

\subsection{2D bulge-disk decomposition in the \textit{r}-band}
\label{sec:Bulge-disk decomposition}
\begin{figure*}
\centering
\includegraphics[width=17cm]{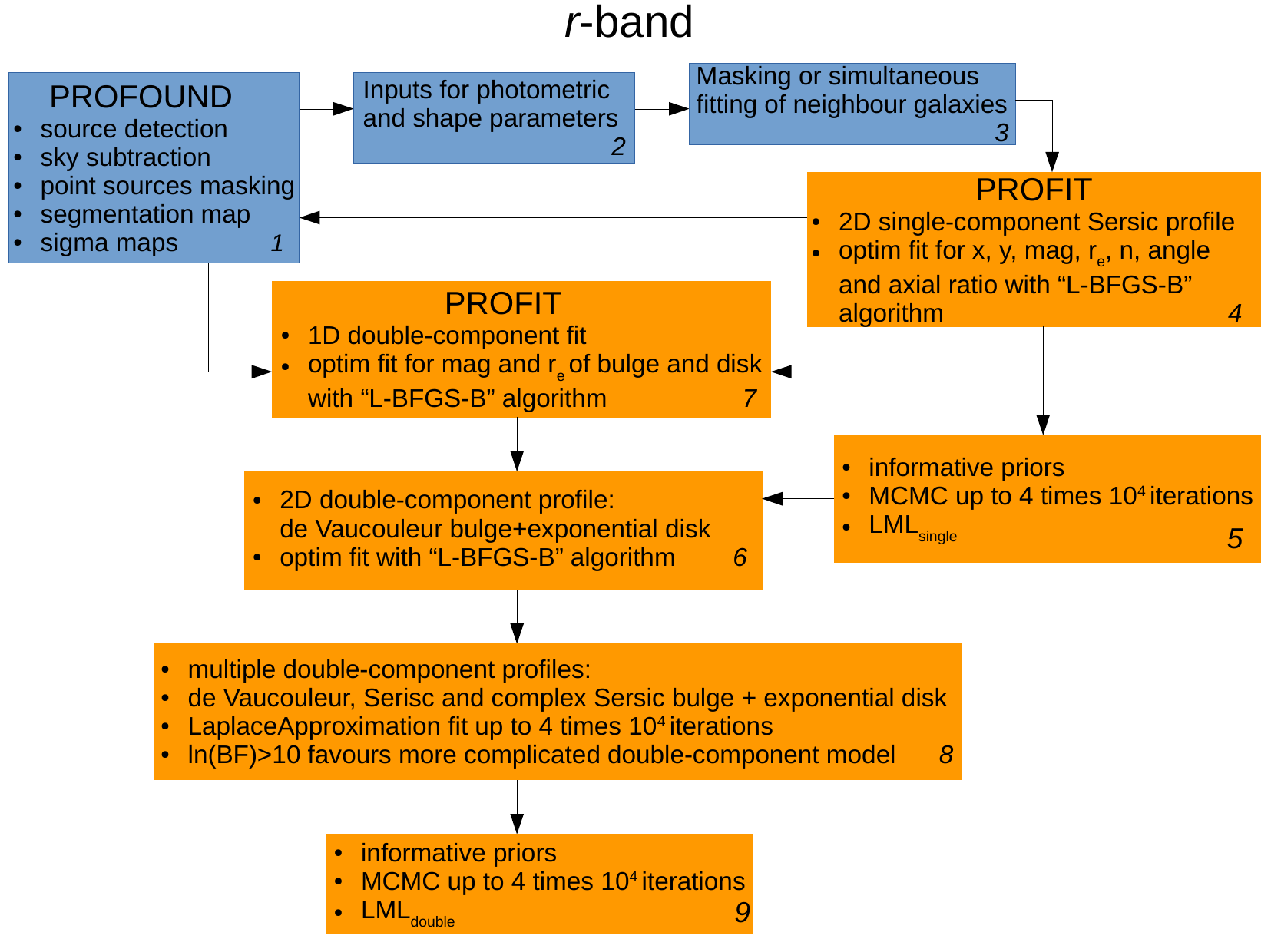}
\caption{Summary of the {\sc ProFound}+{\sc ProFit} pipeline for 2D photometric bulge-disk decomposition in the $r$-band. Blue and orange boxes refer to the {\sc ProFound} and {\sc ProFit} steps, respectively.}
\label{rbandpipeline}
\end{figure*}

The 2D photometric bulge-disk decomposition is based on an iterative {\sc ProFound}+{\sc ProFit} pipeline, fitting the single S\'ersic model and multiple double-component bulge+disk models for each galaxy in the $r$-band. The main steps of this pipeline are summarised in Figure~\ref{rbandpipeline}. The boxes containing the steps of Figure~\ref{rbandpipeline} are numbered and highlighted in the following text with square brackets. 

\subsubsection{Initial processing}
\label{Initial processing}
For each cluster galaxy a 400$^{\prime \prime}\times400^{\prime \prime}$ cutout image is obtained in each band. The {\sc ProFound}+{\sc ProFit} pipeline for bulge-disk decomposition is run on the $r$-band image. We use {\sc ProFound}, which performs source detection and segmentation on our images [box 1 of Figure~\ref{rbandpipeline}]. Photometric and shape parameters are estimated for each detected segment, such as flux weighted $x$ and $y$ centers, magnitude, mean surface brightnesses and elliptical semi-major axes containing 50\% (R50), 90\% (R90) and 100\% (R100) of the flux, concentration and axial ratio [box 2 of Figure~\ref{rbandpipeline}]. 

 We estimate analytical Point-Spread-Functions (PSFs) for each $gri$-band cutout image. We select up to 10 stars based on their values of mean surface brightness containing 90\% of the flux and full-width-at-half-maximum (FWHM). In order to avoid contamination from extended objects, we choose only sources with axial ratio $\geq0.7$. Each star is fitted with a single Moffat profile available in {\sc ProFit} \citep{Moffat1969,Robotham2017}. The results from {\sc ProFound} are used as inputs for the $x$ and $y$ centers, magnitude, FWHM, concentration and axial ratio. For each cutout image we estimate the medians of all star parameter results to generate the 2D PSF. The galaxy model is convolved with the 2D PSF model during the fitting with {\sc ProFit}. The average PSF FWHM for the VST/ATLAS $g$-, $r$- and $i$-band is 1.04$\pm$0.31, 0.84$\pm$0.23 and 0.65$\pm$0.24 arcsec, respectively. These results agree within the 1$\sigma$ interval with those found by \citet{Shanks2015} of 0.95, 0.90 and 0.81 arcsec for the VST $g$-, $r$- and $i$-bands. The average PSF FWHM for the SDSS $g$-, $r$- and $i$-band is 1.12$\pm$0.20, 0.95$\pm$0.22 and 0.88$\pm$0.23 arcsec, respectively. These results agree within the 2$\sigma$ level with those found by \citet{Meert2016} of 1.47, 1.35 and 1.28 arcsec for the SDSS $g$-, $r$- and $i$-bands.

\begin{figure}
\includegraphics[width=\columnwidth]{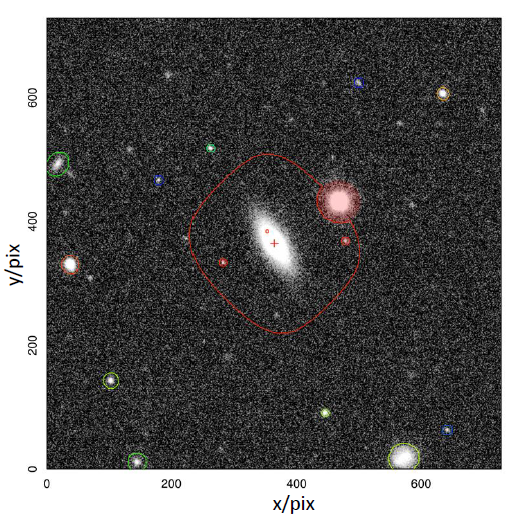}
\caption{$r$-band photometric image for the 9091700038 galaxy, marked by the red central cross. The red contour surrounding the cross shows the segmentation map from {\sc ProFound}. The red color represents the brightest sources, via green, through to the blue color showing the faintest objects. Masked point sources are covered by a red filled circle.}
\label{BDdata}
\end{figure}

Subsequently, we perform source detection and sky subtraction, following the approach of \citet{Owers2019}. We use a stacked $griz$-band SDSS image or $gri$-band VST/ATLAS image (since the $z$-band is not available) as the galaxy data to be analyzed with {\sc ProFound} and then fitted with {\sc ProFit}. The $r$-band and the stacked images are corrected by masking star-like objects that can contaminate the flux of the galaxy of interest. We estimate the sky for each band image and we subtract it individually from each band image before building the stacked image (for a detailed description of the sky estimation see \citealp{Owers2019}). We use {\sc ProFound} to obtain segmentation maps for the corrected $r$-band and stacked images. The segmentation map on the stacked images is used for the initial shape values of the galaxy, while the $r$-band segmentation map gives the initial magnitude value. Pixel-matched sigma maps in the $r$-band are built to propagate the uncertainties. Figure~\ref{BDdata} shows the $r$-band galaxy image with the obtained segmentation map for the 9091700038 galaxy marked by the red cross. Finally, {\sc ProFound} offers all the key inputs for {\sc ProFit}.

\subsubsection{Single-component S\'ersic fits}
\label{Single component fit}
 The single-component S\'ersic profile is used to identify those galaxies that are either bulge- or disk-dominated. We use {\sc ProFit} to fit the primary galaxy and other possible neighbour galaxies with a single-component S\'ersic fit, following \citet{Owers2019}. 

\begin{figure*}
\centering
\includegraphics[width=16cm]{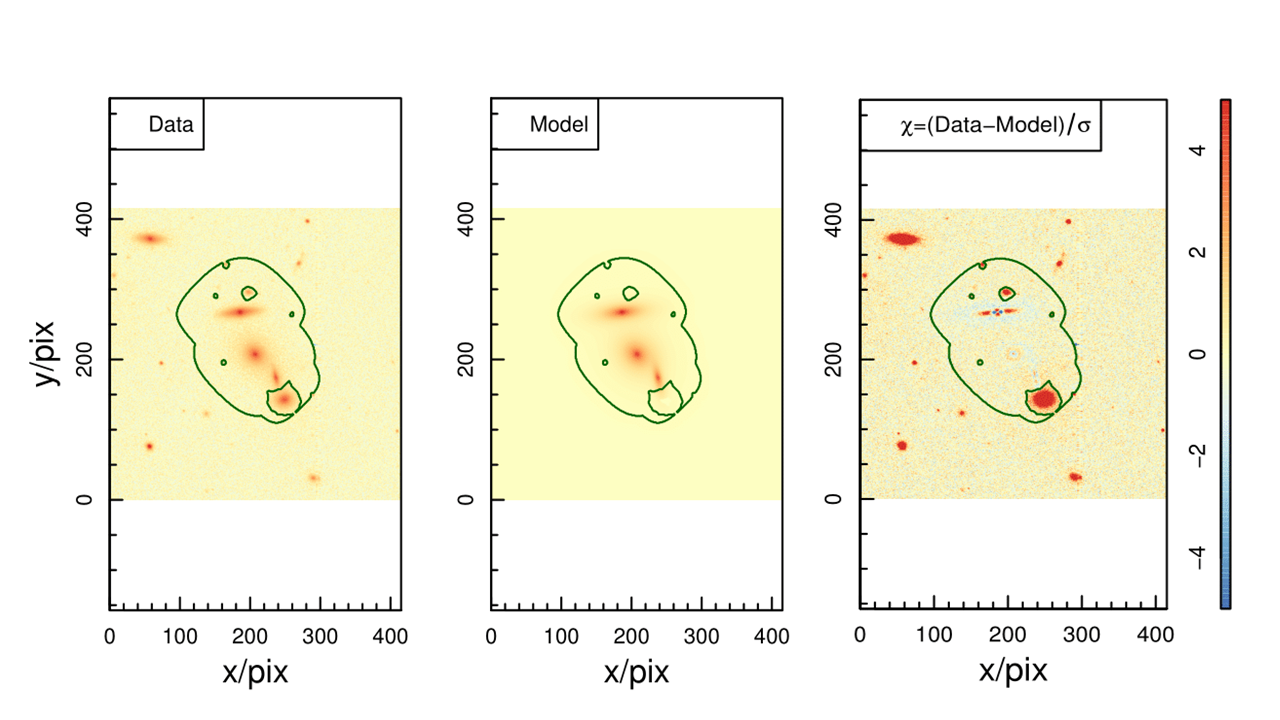}
\caption{Example of 2D {\sc ProFit} plot for three galaxies simultaneously fitted with a single-component S\'ersic model according to the criteria of \citet{Meert2015}. The primary galaxy is the middle one. The data (\textit{left}), model (\textit{middle}) and residuals divided by the sigma map (\textit{right}) are shown. The green line is the segmentation map from {\sc ProFound}. The model plot shows that each galaxy is reproduced, limiting the effects from the neighbours on the primary galaxy fit. Sources having R100 radii that do not overlap with that of the primary galaxy are not fitted.} 
\label{simfitting}
\end{figure*}

Since this work focuses on clusters, there are often near neighbour galaxies that can affect the fits performed on the galaxy of interest. These neighbouring galaxies need to be masked or simultaneously fitted [box 3 of Figure~\ref{rbandpipeline}]. Following the approach of \citet{Meert2015}, we simultaneously fit the primary and the neighbour galaxies when the following criteria are satisfied: (i) the two sources' R100 are overlapping and (ii) the isophotal area of the neighbour is 5\% greater than that of the primary galaxy. If these conditions are not satisfied, the areas occupied by the neighbour sources are masked. Figure~\ref{simfitting} shows an example of the 2D {\sc ProFit} plot for galaxies simultaneously fitted. 

We fit for $x$ and $y$ positions, magnitude, effective radius ($r_{e}$), S\'ersic index ($n$), position angle and axial ratio [box 4 of Figure~\ref{rbandpipeline}]. The fitting is performed using the \textit{R optim} function with the ``L-BFGS-B'' algorithm, since it is a robust, fast algorithm allowing boundaries for the parameters \citep{Broy70,Flet70,Gold70,Shan70}. Table~\ref{fit_limits} shows the adopted boundaries for each fitted parameter. After this first fitting, we subtract the model of the primary galaxy and we run {\sc ProFound} again [boxes 1 and 2 of Figure~\ref{rbandpipeline}]. We check for additional sources to be masked or simultaneously fitted, but missed in the initial segmentation process. We update the segmentation maps and we fit the single S\'ersic model again [box 3 of Figure~\ref{rbandpipeline}].

Following \citet{Moffett2019}, the results of this last optimization are used as initial estimates to the \textit{LaplacesDemon} package (available on github.com/LaplacesDemonR). We use the Markov Chain Monte Carlo (MCMC) LaplacesDemon function with the Component Hit-And-Run Adaptive Metropolis (CHARM) algorithm [box 5 of Figure~\ref{rbandpipeline}]. We set up Gaussian prior distributions for the fitted parameters. We compute the most likely single S\'ersic model over at least $10^{4}$ iterations. In case the outputs from \textit{R optim} are pegged at the lowest or highest allowed fit limits, we use the last {\sc ProFound} values as inputs to MCMC. If convergence is not achieved after $10^{4}$ iterations, we repeat up to three times $10^{4}$ iterations using the MCMC results as inputs. If convergence occurs, we use the mean values from the MCMC stationary samples as parameter estimates. Errors are estimated from the standard deviations of the posterior distributions. Finally, the LaplacesDemon function gives an approximation of the logarithm of the marginal likelihood (LML) that is used to compute the Bayes Factor and to perform model selection. The \textit{top panels} of Figure~\ref{photosummary} show for the 9091700038 galaxy the data (panel \textit{A}), the single-component S\'ersic model (panel \textit{B}), the residuals (panel \textit{C}) and the 1D surface brightness radial profile (panel \textit{D}). The reduced $\chi^2_{\nu}$=1.12, where $\nu$ are the degrees-of-freedom, and LML$_{single}$=$-$78701 of the fit are also reported.

\begin{table*}
 \centering
  \caption{2D bulge-disk decomposition: boundaries for fitted parameters assuming S\'ersic models.} 
  \label{fit_limits}
  \begin{tabular}{@{}lccc@{}}
  \hline
Parameter     &     Single      & Bulge  & Disk\\
\hline

SDSS $x,y$ (pixels) & [$xy_{in}-$5; $xy_{in}$+5]& [$xy_{in}-$5; $xy_{in}$+5]& [$xy_{in}-$5; $xy_{in}$+5]\\
VST/ATLAS $x,y$ (pixels)  & [$xy_{in}-$10; $xy_{in}$+10]& [$xy_{in}-$10; $xy_{in}$+10]& [$xy_{in}-$10; $xy_{in}$+10] \\
mag&[mag$_{in}-$2; mag$_{in}$+2]&[10; 30]&[10; 30]\\
$r_{e}$ (pixels)&[0.1; 10$r_{e,in}$]&[0.1; 2$r_{e,in}$]&[0.1; image size/3]\\
$n$&[0.01; 20]&[0.1; 15]&1\\
position angle&[-180; 360]&[-180; 360]&[-180; 360]\\
axial ratio&[0.0001; 1]&[0.0001; 1]&[0.0001; 1]\\
\hline
\end{tabular}
\end{table*}

\newpage
\subsubsection{Multiple double-component fits}
\label{sec:Multiple double-component fits}
The double-component fitting process adds another component to the single model. We fix the disk model to an exponential profile, while for the bulge model we consider three different profiles with increasing complexity. We perform multiple double-component fits only in the $r$-band, assigning to each galaxy the model characterized by the highest likelihood. For the $g$- and $i$-bands we fit the most likely double-component model assigned in the $r$-band (see Section~\ref{sec:Extension to the $g$ and $i$-bands}).

At first, we fit the double-component de Vaucouleurs + exponential profile [box 6 of Figure~\ref{rbandpipeline}]. The parameters that we fit for are: $x$ and $y$ centers for bulge tying the positions of bulge and disk together, $r$-band magnitude for both components, $r_{e}$ for both components, position angle and axial ratio for disk. We fix the angle and axial ratio of the bulge to 0 and 1, respectively, and the $n$ of the bulge and disk to 4 and 1, respectively. The fitting is performed using the \textit{R optim} function with the ``L-BFGS-B'' algorithm. Table~\ref{fit_limits} shows the adopted boundaries for the fitted parameters.

In order to choose the inputs for the de Vaucouleurs + exponential fit, we follow two approaches. We perform an \textit{R optim} fit for both of the options outlined below, and estimate the likelihood for each approach. The parameters associated with the approach that returns the highest likelihood are used as initial inputs for the de Vaucouleurs + exponential fit. The two approaches are:

\begin{enumerate}
\item We use only the MCMC results of the single S\'ersic fit, where $r_{e,disk}$ is the single S\'ersic fit MCMC output and $r_{e,bulge}=r_{e,disk}/4$. For the initial magnitudes, the flux bulge:disk ratio is 50\%:50\% of the total galaxy flux if the single S\'ersic $n>2$, and 20\%:80\% if $n<2$ [box 5 of Figure~\ref{rbandpipeline}].
\item We combine the MCMC results of the single S\'ersic fit with a 1D de Vaucouleurs + exponential profile. Multiple isophotal ellipses are fitted using the {\sc ProFound} function on the $r$-band galaxy image, building the 1D radial surface brightness profile. We fit for magnitude and $r_{e}$ of both components fixing the $n$ of the bulge and the disk to 4 and 1, respectively. Following the approach of \citet{Lange2016}, we build a grid of 44 inputs for the two components starting from the {\sc ProFound} magnitude and $r_{e}$ values. The more compact component is defined as the bulge and if $r_{e,bulge}$ is more than 10\% larger than $r_{e,disk}$ we swap the radii. We screen the 44 fits excluding solutions pegged at the fitting boundaries. Finally, we take the medians of the results from the remaining 1D double-component fits as the initial inputs for the de Vaucouleurs + exponential profile [box 7 of Figure~\ref{rbandpipeline}]. 
\end{enumerate}

The \textit{R optim} results of the de Vaucouleurs + exponential profile are used as inputs to the \textit{LaplacesDemon} package, which is used in LaplaceApproximation mode with the ``L-BFGS-B'' algorithm to perform multiple double-component fits [box 8 of Figure~\ref{rbandpipeline}]. If the \textit{R optim} results are pegged at the fitting boundaries or the output $r_{e,bulge}$ is more than 10\% larger than $r_{e,disk}$, the inputs are derived from the most likely initial 2D model. To determine the level of complexity in the double-component fits that are allowed by the data, we perform fits for the following double-component varieties, listed in order of increasing complexity:
\begin{enumerate}
\item de Vaucouleurs + exponential;
\item simple S\'ersic + exponential, where $n$ of the bulge is free;
\item S\'ersic + exponential, where $n$, axial ratio and position angle of the bulge are free.
\end{enumerate}

The likelihood of each profile is maximized over $10^{4}$ iterations up to four times if convergence is not achieved. We set up Gaussian priors for the fitted parameters and the following constraint to avoid unphysical solutions: if $r_{e,bulge}>r_{e,disk}$, we re-initialize the value of $r_{e,bulge}$ to the value of $r_{e,disk}$. For each of the double-component models, the LaplaceApproximation function estimates the logarithm of the marginal likelihood (LML) and it is less time consuming than the LaplacesDemon function. LML is subsequently used for model selection using the Bayes Factor (BF). We choose the more complex model if $\ln$(BF)$>10$ in order to avoid over fitting. For most galaxies the S\'ersic + exponential is the most likely double-component model.

\begin{figure*}
\includegraphics[width=19cm]{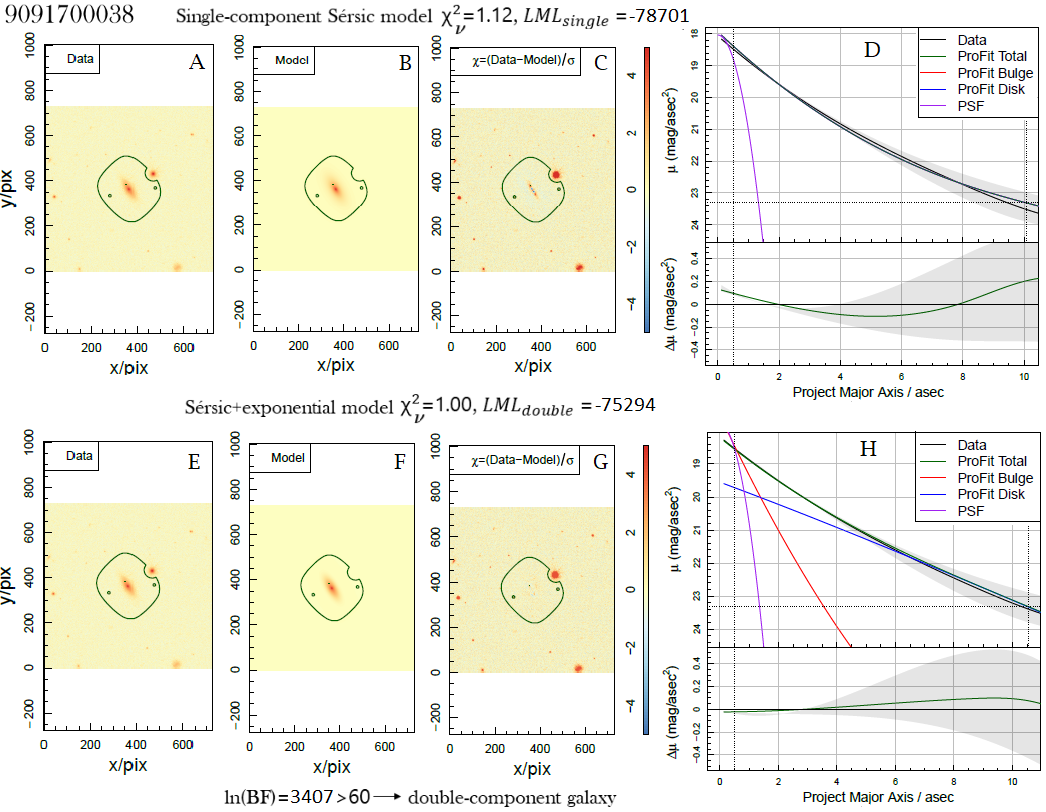}
\caption{Summary plot for the 2D photometric bulge-disk decomposition for the 9091700038 galaxy. The \textit{A to D panels} and the \textit{E to H panels} represent the {\sc ProFit} results for the single-component S\'ersic fit and double-component S\'ersic + exponential fit, respectively. The data (\textit{A,E}), the model (\textit{B,F}), the residuals divided by the sigma map (\textit{C,G}) and the 1D surface brightness radial profiles with associated uncertainty (\textit{D,H}) are shown. The green contour shows the segmentation map from {\sc ProFound}. The statistical parameters $\chi^2_{\nu}$, LML and $\ln$(BF) are reported, showing that the galaxy is classified as double-component.}
\label{photosummary}
\end{figure*}

The parameters of the best model as determined by the LaplaceApproximation approach are used as initial inputs to the full MCMC run. We again use the LaplacesDemon function with the CHARM adaptive algorithm and $10^{4}$ iterations up to four times [box 9 of Figure~\ref{rbandpipeline}]. If the LaplaceApproximation results are pegged at the fitting boundaries or the output $r_{e,bulge}$ is more than 10\% larger than $r_{e,disk}$, we choose as inputs to MCMC the results from \textit{R optim}. We update the prior distributions around the MCMC inputs and we re-initialize the value of $r_{e,bulge}$ to the value of $r_{e,disk}$ when $r_{e,bulge}>r_{e,disk}$. Finally, if convergence occurs, we use the mean values from the MCMC stationary samples as parameter estimates. The errors are estimated from the standard deviations of the posteriors distributions and an estimate of the final LML for the chosen double-component model is given. The \textit{bottom panels} of Figure~\ref{photosummary} show for the 9091700038 galaxy the data (panel \textit{E}), the double-component S\'ersic + exponential model (panel \textit{F}), the residuals (panel \textit{G}) and the 1D surface brightness radial profile (panel \textit{H}). The reduced $\chi^2_{\nu}$=1.00 and LML$_{double}$=$-$75294 of the fit are also reported.

\subsection{Extension to the \textit{g}- and \textit{i}-bands}
\label{sec:Extension to the $g$ and $i$-bands}
The bulge-disk decomposition is performed in the $g$- and $i$-band images using a similar {\sc ProFound}+{\sc ProFit} fitting pipeline to the one for the $r$-band images described in Section~\ref{sec:Bulge-disk decomposition}. The {\sc ProFound} part for the stacked images is the same as for the $r$-band in Section~\ref{Initial processing}, but the band-dependent quantities such as the image, the PSF, the sky and the segmentation map come from the selected band. We fit with {\sc ProFit} a single S\'ersic profile using for photometric inputs the {\sc ProFound} values derived from the respective band segmentation map and for shape inputs the $r$-band single S\'ersic results obtained in Section~\ref{Single component fit}. Then, we fit a double-component profile choosing the model that has been selected in the $r$-band in Section~\ref{sec:Multiple double-component fits}. For both the single/double-component fits we fix the position angle and axial ratio to the $r$-band results, while we fit for position, magnitude, effective radius and S\'ersic index. Finally, we follow the same approach of optimization, MCMC implementation and LML estimation previously implemented for the $r$-band images.

\subsection{Characterizing single- and double-component galaxies}
\label{sec:Model selection}
We perform the 2D bulge-disk decomposition for 1204 SDSS and 591 VST/ATLAS cluster galaxies, as selected in Section~\ref{sec:Data}. The pipelines {\sc ProFound}+{\sc ProFit} work in the $r$-, $g$- and $i$-bands for 1730 galaxies. The model selection is performed in the $r$-band. Both the single-/double-component fits converge in the $r$-band for 1126 SDSS and 529 VST/ATLAS galaxies, i.e. for $\sim$96\% of the galaxy sample. To select the most likely model justified by the data we use the Bayes Factor $\ln(\rm{BF})=\rm{LML_{double}-LML_{single}}$. Generally, $\ln(\rm{BF})>5$ is considered as strong evidence favouring the more complicated model \citep{Kass1995,Gordon2017}. The visual inspection of our fits led us to use the more stringent criterion $\ln(\rm{BF})>60$, which ensures that we avoid contamination by single-component galaxies. 

\begin{figure*}
\includegraphics[width=19cm]{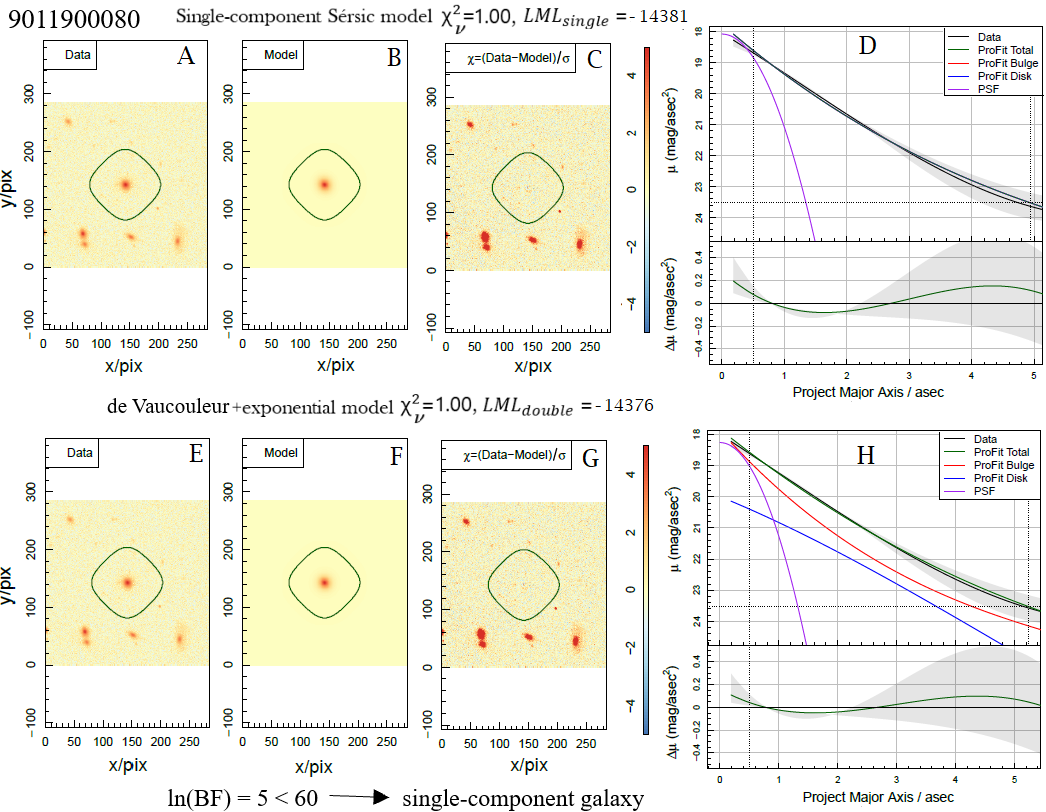}
\caption{Summary plot for the 2D photometric bulge-disk decomposition for the 9011900080 galaxy. The \textit{A to D panels} and the \textit{E to H panels} represent the {\sc ProFit} results for the single-component S\'ersic fit and double-component de Vaucouleurs + exponential fit, respectively. The data (\textit{A,E}), the model (\textit{B,F}), the residuals divided by the sigma map (\textit{C,G}) and the 1D surface brightness radial profiles with associated uncertainty (\textit{D,H}) are shown. The same $\chi^{2}_{\nu}$ values, the same residuals and the 1D surface brightness profiles for the single-component fit and the double-component fit suggest that the galaxy is single-component, justifying our more stringent criterion $\ln(\rm{BF})>60$ to select double-component galaxies.}
\label{photosummarySingle}
\end{figure*}

However, the inclusion of galaxies with $\ln(\rm{BF})>5$ does not affect our final results (see Section~\ref{Trends with galaxy density}). We find 1083 double-component galaxies with $\ln(\rm{BF})>60$, while 572 are selected as single-component galaxies having $\ln(\rm{BF})<60$. The 9091700038 example galaxy in Figure~\ref{photosummary} is double-component with $\ln(\rm{BF})=3407$. For comparison, Figure~\ref{photosummarySingle} shows an example of single-component galaxy with $\ln(\rm{BF})=5$. The same $\chi^{2}_{\nu}$ values and the same residuals for the single-component fit and the double-component fit justify our more stringent criterion $\ln(\rm{BF})>60$ to select double-component galaxies. The 1D surface brightness profile for the double-component fit is dominated by the bulge light, suggesting that the single-component fit is better justified for this galaxy.
 
In addition to the selection using $\ln(\rm{BF})$, we filter our sample of 1083 double-component galaxies. Our aim is to select S0 galaxies for which we can reliably measure the colors of both the bulge and the disk components. To this end, we exclude 68 double-component galaxies with unphysical fits:
\begin{itemize}
\item $n_{bulge}$ pegged at the allowed fit limits (37 galaxies);
\item $r_{e,bulge}$ or $r_{e,disk}$ pegged at the allowed fit limits (31 galaxies).
\end{itemize}
We also exclude 404 double-component galaxies with unresolved radial parameters:
\begin{itemize}
\item $r_{e,disk}$ is less than 5\% larger than $r_{e,bulge}$, since the bulge is defined as the most compact component (134 galaxies);
\item $r_{e,bulge}$ is smaller than 80\% of the PSF half-width-at-half-maximum, in order to recover reliable bulge properties (\citealp{Gadotti2009,Meert2015}; 270 galaxies).
\end{itemize}
Finally, we exclude 142 double-component galaxies for which the separation of the bulge and disk measurements is unreliable:
\begin{itemize}
\item $n_{bulge}<0.6$ to exclude bars, which typically have $n\sim0.5$ (\citealp{Gadotti2008,Gadotti2008b}; 33 galaxies);
\item bulge-to-total flux ratio lower than 0.2 and larger than 0.8 to exclude galaxies dominated by one component (\citealp{Meert2015}; 109 galaxies). 
\end{itemize}
The filter causing the exclusion of the largest number of double-component fits is due to impact of the PSF on the geometrical properties of the bulge, which tends to make the bulge rounder and harder to constrain. The effects of the PSF on the bulge properties have been detected in several previous works \citep{Gadotti2008,Gadotti2009,Bernardi2014,Meert2015}. For the galaxies with double-component unphysical fits or with unreliable bulge/disk separation, the single-component fit is preferable. The galaxies with unresolved radial parameters are not useful for investigating separate bulge and disk properties, but they are still better characterized by a double-component fit. In conclusion, the number of SDSS+ATLAS single-component galaxies is 782 (47\%), the number of SDSS+ATLAS reliable double-component galaxies is 469 (28\%) and the number of SDSS+ATLAS double-component galaxies with unresolved radial parameters is 404 (25\%). For the single-component sample, 388 galaxies are bulge-dominated with S\'ersic index $n>2$ and 394 are disk-dominated with S\'ersic index $n<2$. For the reliable double-component sample, 418 galaxies are better described by the most complex S\'ersic + exponential model, 37 by the simple S\'ersic + exponential model and the remaining 14 galaxies by the de Vaucouleurs + exponential model. The range in stellar mass of the 469 double-component galaxies is $9.5<\log(M_{*}/M_{\odot})<11.5$. The galaxies with unresolved bulge and disk radial parameters are characterized by a lower galaxy stellar mass and a smaller size compared to the double-component galaxies with reliable bulge and disk measurements. The median $\log(M_{*}/M_{\odot})$ difference is 0.145$\pm$0.037 dex and the median galaxy size difference, measured as the single-component $r_{e}$, is 0.469$\pm$0.091 arcsec. Thus, the exclusion of these galaxies limits our study to double-component galaxies that tend to have large stellar mass and size, for which the bulge and disk colors are reliably estimated.

\subsection{Internal tests}
\label{Internal tests for 2D bulge-disk decomposition}
In order to determine the reliability of the single- versus double-component characterization from Section~\ref{sec:Model selection}, in this Section we perform a number of internal checks.

\begin{figure*}
\centering
\includegraphics[width=0.45\textwidth]{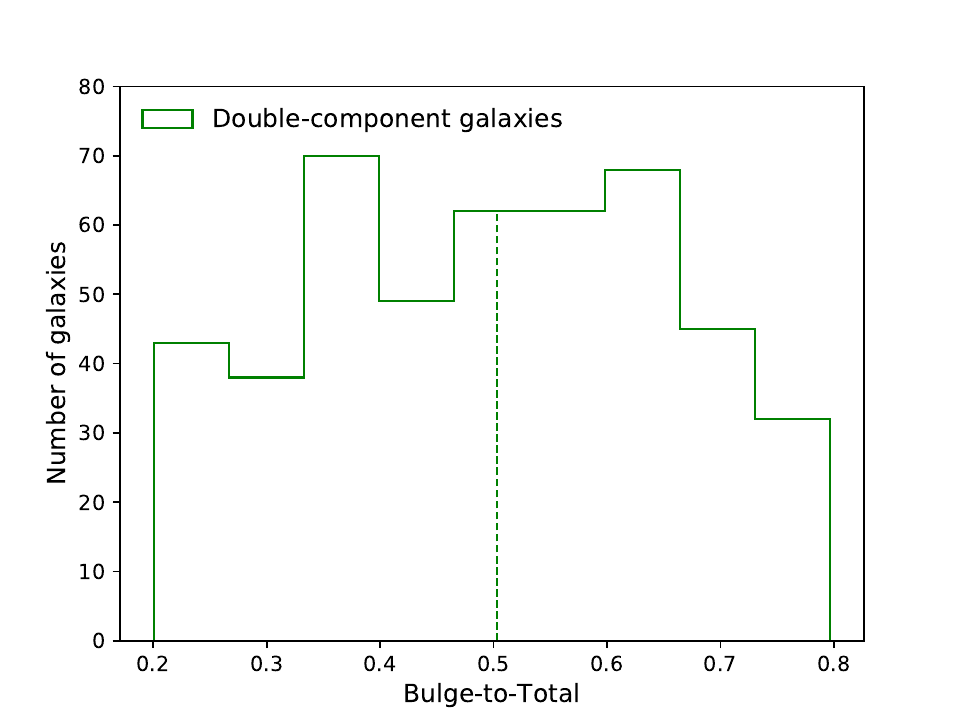}
\includegraphics[width=0.45\textwidth]{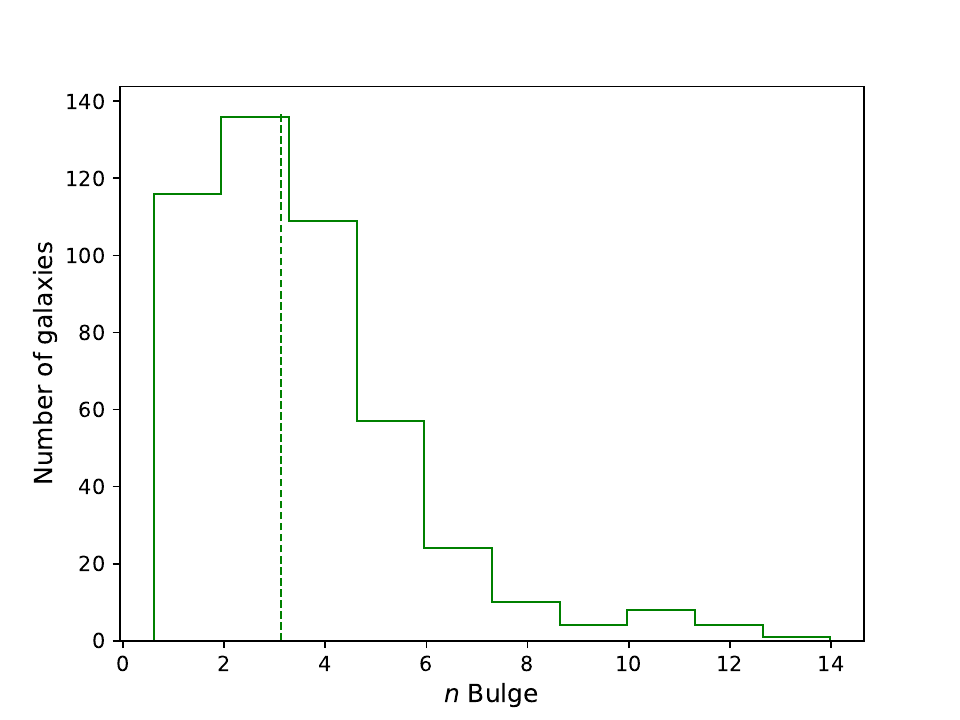}
\caption{Histograms in bulge-to-total (B/T) flux ratio (\textit{left panel}) and $n_{bulge}$ (\textit{right panel}) for 469 SDSS+ATLAS double-component galaxies. The medians at B/T=0.50 and $n_{bulge}$=3.13 show the reliability of the model selection.}
\label{DoubleHistograms}
\end{figure*}

We explore the bulge-to-total flux ratio (B/T) and the S\'ersic index of the bulge ($n_{bulge}$) for the double-component sample. For a typical galaxy with two components we expect $n_{bulge}=4$ and B/T not assuming too high or low values which indicate more bulge-dominated or disk-dominated galaxies. For 469 SDSS+ATLAS double-component galaxies the \textit{left panel} of Figure~\ref{DoubleHistograms} shows the distribution of the B/T values. The distribution is uniform with deficits at the lowest and highest B/T extremes where the data do not justify multiple components. The distribution of the $n_{bulge}$ in the \textit{right panel} of Figure~\ref{DoubleHistograms} peaks between 2 and 4. These outcomes highlight the reliability of the model selection. Similar analyses are conducted by \citet{Meert2015} and \citet{Lange2016}, who performed bulge-disk decomposition on SDSS galaxy images. \citet{Lange2016} found that double-component galaxies have mainly $n_{bulge}=2-4$ (see their Figure 7). \citet{Meert2015} observed that double-component galaxies peaked for $n_{bulge}=1-4$ (see their Figures 8 and 9), and that lowest and highest B/T values are preferred by single-component fits (see their Figure 12). 

\begin{figure*}
\centering
\includegraphics[width=0.45\textwidth]{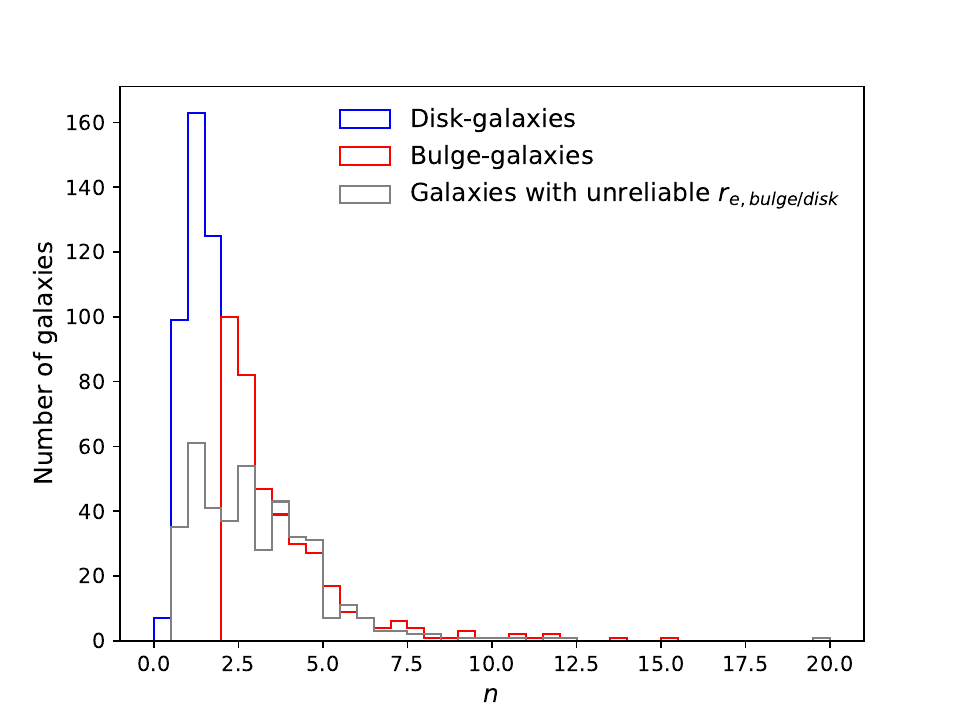}
\includegraphics[width=0.45\textwidth]{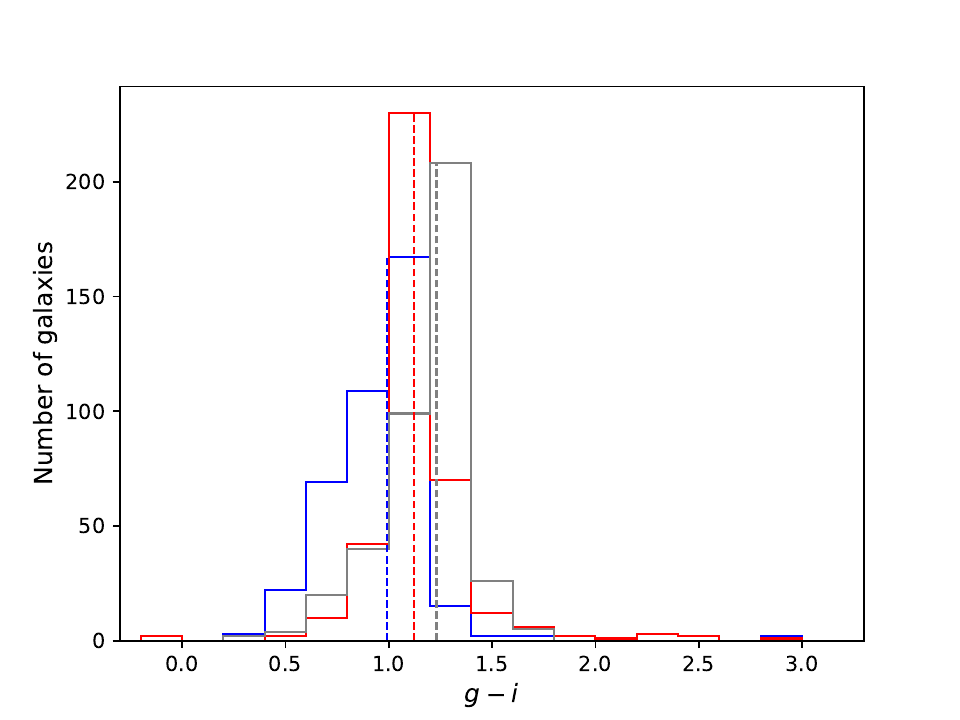}
\caption{Histograms in $n$ (\textit{left panel}) and $g-i$ color (\textit{right panel}) for 782 SDSS+ATLAS single-component galaxies and for 404 SDSS+ATLAS double-component galaxies with unreliable radial parameters. The dashed lines represent the median values. Disk-dominated galaxies (single S\'ersic $n<$2; blue) show bluer color compared to bulge-dominated galaxies (single S\'ersic $n>$2; red). Double-component galaxies with unreliable $r_{e,bulge/disk}$ (grey) are mainly bulge-dominated and red. }
\label{SingleHistograms}
\end{figure*}

In the \textit{left panel} of Figure~\ref{SingleHistograms} we investigate the separate distributions of $n$ for the 782 SDSS+ATLAS single-component galaxies and for the 404 galaxies having double-component fits with unreliable radial parameters. The 782 single-component galaxies are divided into bulge-dominated galaxies with $n>2$ and disk-dominated galaxies with $n<2$. We expect disk-dominated galaxies to be characterized by active star formation and to show bluer colors than bulge-dominated galaxies. The \textit{right panel} in Figure~\ref{SingleHistograms} shows the $g-i$ colors for the bulge-dominated galaxies with $n>2$, disk-dominated  with $n<2$ and double-component galaxies with unreliable $r_{e,bulge/disk}$. The $g$ and $i$ apparent magnitudes have been corrected for the interstellar extinction of the Milky Way using the dust maps of \citet{Schlegel1998} and for the K-correction \citep{Blanton2007}. As expected, the disk-dominated galaxies are characterized by bluer colors when compared with the bulge-dominated systems. Double-component galaxies with unreliable radial parameters tend to be more bulge-dominated with average $n=3.07$. They also show red colors, highlighting an old bulge-dominated population.

\begin{figure*}
\centering
\includegraphics[width=0.45\textwidth]{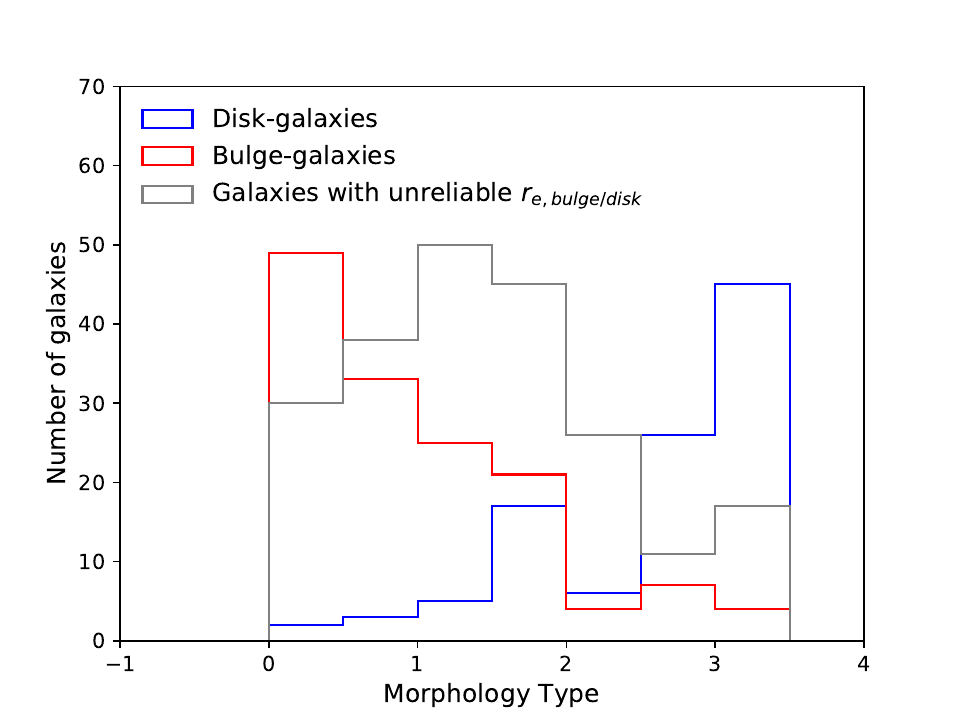}
\includegraphics[width=0.45\textwidth]{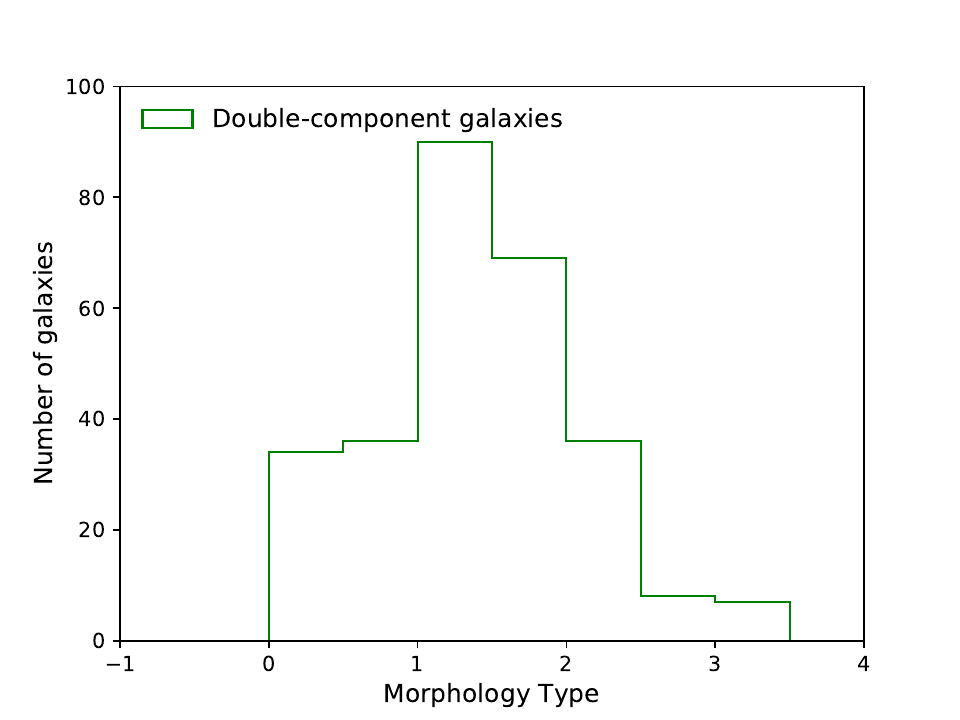}
\caption{Morphology histograms for 247 single-component galaxies divided into bulge-dominated and disk-dominated (\textit{left panel}), for 217 double-component galaxies having unreliable $r_{e,bulge/disk}$ (\textit{left panel}) and for 280 double-component galaxies (\textit{right panel}). Single-component bulge-dominated galaxies are mainly early-type galaxies, while disk-dominated galaxies are late-spirals. Double-component galaxies have a peak for S0 galaxies.}
\label{morphology}
\end{figure*}

We investigate the single/double-component galaxy samples as a function of morphology using the classification performed for the SAMI Galaxy Survey by \citet{Cortese2016}. The cluster portion of the SAMI Galaxy Survey contains a subset of the galaxies in the current sample \citep{Owers2017}, therefore morphological classifications are only available for: 280/469 reliable double-component galaxies, 217/404 double-component galaxies with unresolved radial parameters and 247/782 single-component galaxies. The \textit{left panel} in Figure~\ref{morphology} shows the morphology distributions of single-component galaxies divided into bulge-dominated, disk-dominated and galaxies with double-component fits having unreliable $r_{e,bulge/disk}$. The \textit{right panel} in Figure~\ref{morphology} shows the morphology distribution for the double-component galaxies. The galaxy morphology is represented by: elliptical=0, elliptical/S0=0.5, S0=1, S0/early-spiral=1.5, early-spiral=2, early/late-spiral=2.5 and late-spiral=3. We find that single-component bulge-dominated galaxies are classified as elliptical (34\%), elliptical/S0 (23\%), S0 (17\%), S0/early-spiral (18\%), early/late-spiral (5\%) and late-spiral (3\%). The contamination from late-type galaxies may be due to the inherent difficulty associated with the visual morphological classification. Regarding the bulge-disk decomposition, most of these galaxies show asymmetrical features which complicate the fitting process. Complications in fitting the bulge component might arise from the presence of bars or point sources. Disk-dominated galaxies are mainly late-spiral (43\%), early/late-spiral (25\%) and S0 (22\%). Double-component galaxies with unreliable $r_{e,bulge/disk}$ show a high percentage of S0s, in agreement with the fact that the more complicated model is favored for them having $\ln(\rm{BF})>60$. Reliable double-component galaxies have a peak for S0 galaxies, with $\sim$70\% classified as S0/possible S0. Thus, they represent a consistent sample of visually selected S0 galaxies.

Finally, to test the reliability of the 2D bulge-disk decomposition we perform an internal check for the cluster A85, where there are 226 galaxies with measurements common to both SDSS and VST/ATLAS imaging. Comparing the fits between the two catalogues gives us a quantitative idea of the amount of systematic uncertainty due to the different imaging used in the sample. We find 106 and 113 galaxies with $\ln(\rm{BF})>60$ using the SDSS and VST/ATLAS measurements, respectively. 90 galaxies have $\ln(\rm{BF})>60$ using both the SDSS and VST/ATLAS data. Applying the filters of Section~\ref{sec:Model selection} we find 33 reliable SDSS double-component galaxies, 33 reliable VST/ATLAS double-component galaxies and 18 reliable double-component galaxies in common. Most of the galaxies not in common have unresolved $r_{e,bulge}$ with respect to the PSF FWHM for SDSS galaxies. The relative differences between the SDSS and VST/ATLAS magnitude, axial ratio and position angle measurements of both components are smaller than 2\% with dispersion $\sim3$\%, indicating good agreement between the two imaging surveys for these parameters. Highest offsets of $\sim5-7$\% with dispersions $\sim9-12$\% are observed for the effective radii and the bulge S\'ersic index where the SDSS results are larger compared to the results obtained from the VST/ATLAS data. Similar conclusions are drawn for the fitted parameters in the $g$- and $i$-bands. The large scatter and offset for the bulge measurements reflect the inherent difficulty in measuring the bulge parameters. They also might be driven by the larger PSF FWHM of the SDSS data compared to the PSF FWHM of the VST/ATLAS data.

In agreement with our results, \citet{Owers2019} compared the $r$-band results for the A85 single-component galaxies from the SDSS and VST/ATLAS imaging, observing larger offsets for the effective radius and the S\'ersic index in the SDSS data. They found that the systematic differences in $n$ and $r_{e}$ are mainly observed for bulge-dominated galaxies. This suggests that the systematic offsets are likely due to the over-subtraction of the sky around these larger objects in the VST/ATLAS data. This sky over-subtraction leads to a steepening of the outer proﬁle and to smaller $n$ and $r_{e}$ values with respect to those derived from the SDSS data. 

The internal tests carried for the single- and double-component galaxy samples show the reliability of the galaxy characterization. Single-component galaxies show the typical properties of bulge/disk-dominated galaxies, while the double-component sample justifies the choice of two components to describe the data. Finally, Appendix~\ref{2D bulge-disk decomposition: the catalogues} describes the catalogues obtained from the 2D bulge-disk decomposition, which are fully available in the machine-readable version. We present one catalogue for each band, containing the 1730 SDSS+ATLAS cluster galaxies for which the pipelines {\sc ProFound}+{\sc ProFit} work.

\section{Results}
\label{sec:Results}
We aim to shed light into the formation of S0 galaxies in clusters. In particular, we focus on understanding the relative importance of environmental processes acting in the cluster core and pre-processing acting in the cluster outskirts. We study the $g-i$ colors separately for bulges and disks up to $\sim2.5\,R_{200}$, including the infalling cluster regions where galaxies are mainly in groups or isolated. We explore the color-magnitude relations for the two galaxy components. To understand the impact of the environment, we investigate the dependence of the bulge and disk $g-i$ colors on the projected cluster-centric distance $R/R_{200}$ and on the local galaxy density. We disentangle backsplash and infalling galaxy populations in the projected phase space. Finally, we compare the results for double-component galaxies with single-component disk-dominated galaxies.

\subsection{Color-magnitude relations}
\label{Color-magnitude relations}
We aim to understand whether the bulge, the disk or both components are responsible for the global galaxy color-magnitude relation observed in clusters. The study of the bulge and disk color-magnitude trends help us to assess mechanisms that drove the evolution history of the two components. We explore the extinction- and K-corrected bulge and disk $g-i$ colors for the 469 double-component galaxies within $2.5\,R_{200}$ and with reliable bulge and disk measurements as described in Section~\ref{sec:Model selection}.

\begin{figure}[h!]
\centering
\includegraphics[width=\columnwidth]{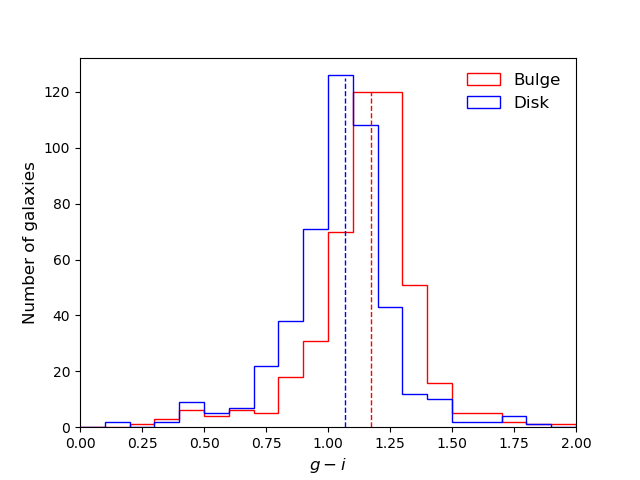}
\caption{Histogram in $g-i$ color for bulges and disks of the 469 double-component galaxies within $2.5\,R_{200}$. The dashed lines represent the median values. Disks are shifted towards bluer colors with respect to bulges by 0.11$\pm$0.02 mag.}
\label{colorsCMRhistogram}
\end{figure}

Figure~\ref{colorsCMRhistogram} shows the $g-i$ color distributions for the bulges and the disks. The disks are mainly shifted towards bluer colors with respect to the bulges. The median $g-i$ offset separating the two distributions is 0.11$\pm$0.02 mag. We find that 67$\pm$2\% of the galaxies have a bulge that is redder than the disk, while for the remaining 33$\pm$2\% the bulge is bluer than the disk. 

\begin{figure*}
\centering
\includegraphics[width=0.45\textwidth]{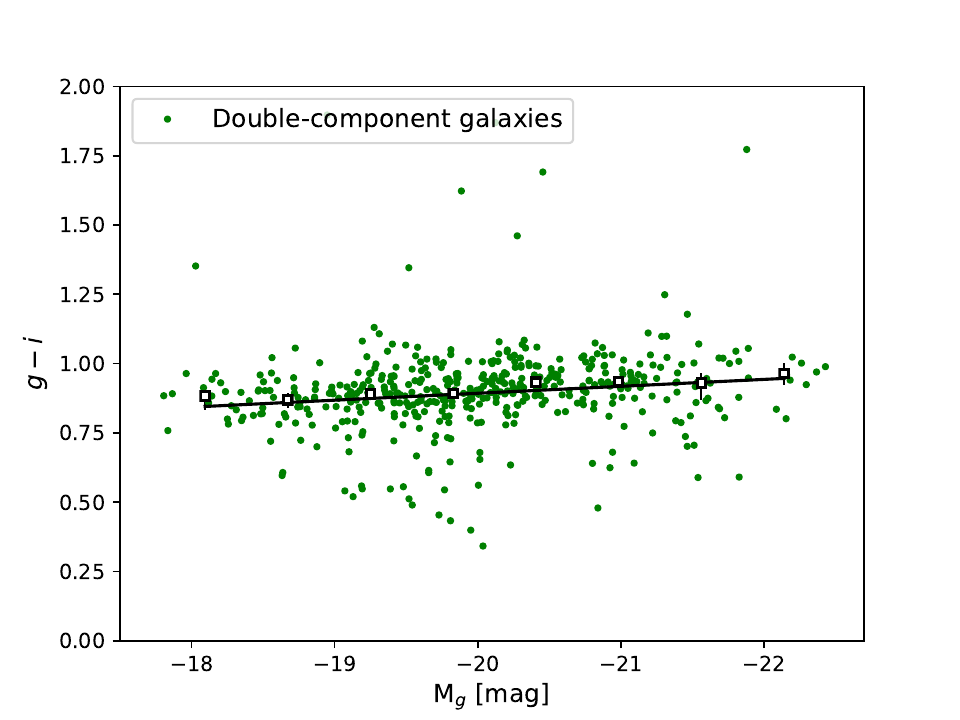}
\includegraphics[width=0.45\textwidth]{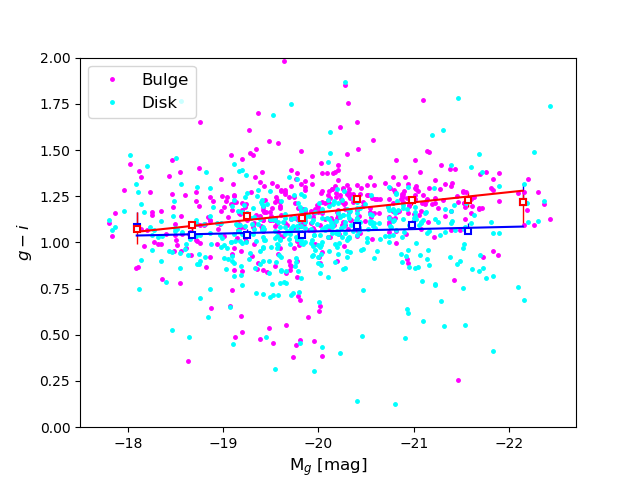}
\caption{Color-magnitude diagrams for the 469 double-component galaxies within $2.5\,R_{200}$ (\textit{left panel}) and for the bulges (red, magenta) and the disks (blue, light blue) separately (\textit{right panel}). In both panels $M_{g}$ is the total absolute magnitude of the galaxy. Best-fitting lines are estimated with the Hyper-Fit software \citep{Robotham2015}. Squares represent the median values in magnitude bins with the respective robust standard errors. Bulges show a statistically significant color-magnitude relation ($4.6\sigma$ level), while for disks the slope is consistent with zero.}
\label{colorsCMR}
\end{figure*}

We investigate the $g-i$ color-magnitude relations for double-component galaxies and separately for bulges and disks in the \textit{left} and \textit{right panels} of Figure~\ref{colorsCMR}, respectively. The total absolute magnitude of the galaxy $M_{g}$ has been estimated using $M_{g}=g-5\log_{10}(D_{L}/10)$ where $D_{L}$ is the luminosity distance measured using the cluster redshift. K-correction and Milky Way extinction corrections are applied. We fit the $g-i$ color-magnitude values using the Hyper-Fit software which takes into account the random uncertainties associated with the data \citep{Robotham2015}.

The slope of the best-fitting line for double-component galaxies (in black) is $-$0.025$\pm$0.008, implying a statistically significant color-magnitude relation at the $3\sigma$ level. We test for the color-magnitude dependency with a Spearman rank correlation test, finding the correlation coefficient $\rho$=$-$0.26 and P-value=6.22$\times10^{-9}$. P-values smaller than 0.05 indicate that we can reject the null hypothesis of uncorrelated data. These results confirm a color-magnitude correlation for double-component galaxies. The best-fitting line for bulges (in red) has a slope equal to $-$0.055$\pm$0.012 and the slope for disks (in blue) is $-$0.012$\pm$0.014. The trend for bulges is statistically significant at the $4.6\sigma$ level, while for disks it differs from a zero slope by $<1\sigma$. The results of the Spearman rank correlation test are $\rho$=$-$0.29 and P-value=1.29$\times10^{-10}$ for bulges, $\rho$=$-$0.08 and P-value=0.094 for disks. Thus, the bulge component primarily contributes to the global color-magnitude relation within $\sim 2.5\,R_{200}$.

\begin{figure*}
\centering
\includegraphics[width=0.45\textwidth]{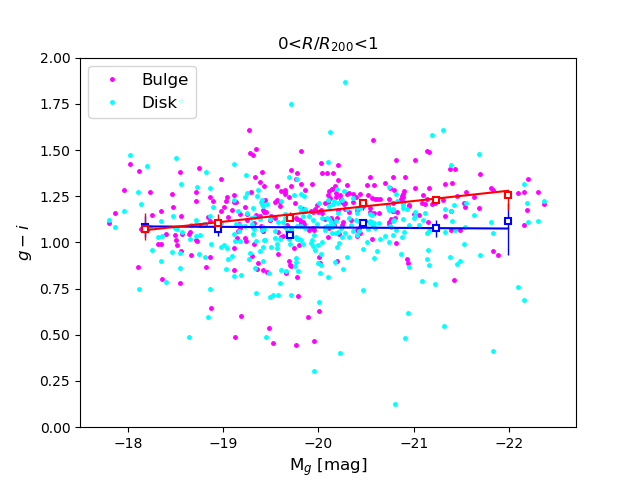}
\includegraphics[width=0.45\textwidth]{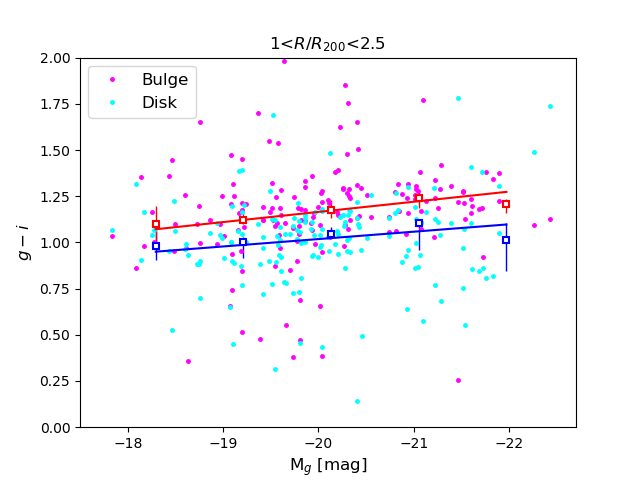}
\caption{Bulge/disk color-magnitude diagrams for the virialized galaxy population within $R<1\,R_{200}$ (\textit{left plot}) and for the infalling/backsplash galaxy population within $1\leq R/R_{200}<2.5$ (\textit{right plot}) samples. Color-magnitude correlations are found for bulges within and beyond $1\,R_{200}$. The disk color-magnitude relations are flat within and beyond $1\,R_{200}$.}
\label{colorsCMRradial}
\end{figure*}

We explore the bulge/disk color-magnitude relations separately for galaxies within $1\,R_{200}$, where a large fraction of galaxies are more likely to have spent an extended period of time (i.e., the virialized galaxy population), and for galaxies with $1\leq R/R_{200}<2.5$, which have been recently accreted or have already traversed the cluster core and are observed close to the maximum distance before their second infall (i.e., the mixed infalling/backsplash galaxy population). For both the virialized and infalling/backsplash galaxies, $\sim70$\% of the bulges are redder than the disks. Figure~\ref{colorsCMRradial} displays the bulge/disk color-magnitude relations for the two sub-samples. Table~\ref{CMslopes} summarizes the slope values of the bulge/disk color-magnitude relations, the Spearman rank correlation coefficients and the associated P-values for the total sample and the two sub-samples. Bulges show a steep color-magnitude relation for the virialized galaxy population that differs from a zero slope at the $\sim4.3\sigma$ level. P-values$<0.05$ indicate correlation between the bulge $g-i$ colors and the total $g$-band absolute magnitude of the galaxies. A similar trend is observed for the bulges of the mixed infalling/backsplash sample. The slopes of the color-magnitude trends for the disks are consistent with zero for both the inner and outer radial regions. P-values$>0.05$ confirm the null hypothesis of uncorrelated color-magnitude data for the disks.

\begin{table*}
 \centering
  \caption{Color-Magnitude relations as a function of $R/R_{200}$. Column 1 lists the sample, column 2 the radial distance, column 3 the number of galaxies, columns 4/5 the bulge/disk slopes, columns 6/7 the Spearman rank correlation coefficients with the associated P-values for bulges, and columns 8/9 the Spearman coefficients and P-values for disks.} 
  \label{CMslopes}
  \begin{tabular}{@{}lcccccccc@{}}
  \hline
Galaxy Sample     &     $R/R_{200}$  & $N_{gal}$& $CM_{bulge}$ & $CM_{disk}$ & $(CM)\rho_{bulge}$ &$(CM)P_{bulge}$ & $(CM)\rho_{disk}$ &$(CM)P_{disk}$\\
\hline
Total  & 0.0 $-$ 2.5 & 469 &  $-$0.055$\pm$0.012&$-$0.012$\pm$0.014&$-$0.29&1.29$\times10^{-10}$&$-$0.08&0.094\\
Virialized  & 0.0 $-$ 1.0&  300&  $-$0.056$\pm$0.013& \hspace{3mm}0.003$\pm$0.017&$-$0.29&2.27$\times10^{-7}$&$-$0.05&0.410\\
Infalling/Backsplash  & 1.0 $-$ 2.5&169 &  $-$0.055$\pm$0.020&$-$0.040$\pm$0.022&$-$0.28&1.97$\times10^{-4}$&$-$0.12&0.117\\
\hline
\end{tabular}
\end{table*}

\begin{figure*}
\centering
\includegraphics[width=0.45\textwidth]{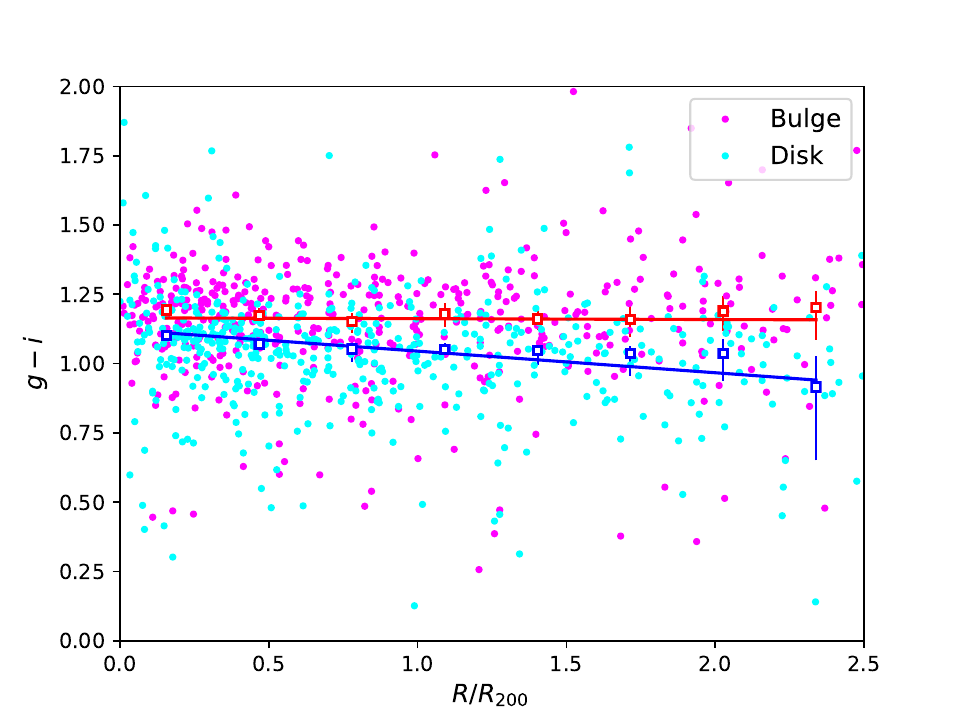}
\includegraphics[width=0.45\textwidth]{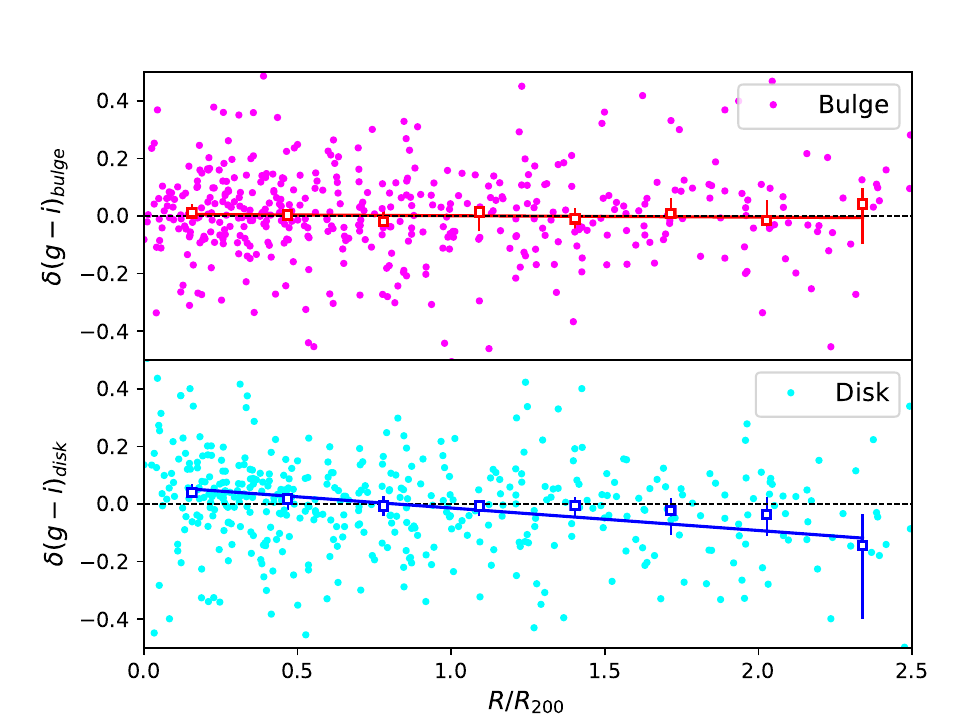}
\caption{$g-i$ colors (\textit{left panel}) and $g-i$ colors decoupled from the color-magnitude relations (\textit{right panel}) versus $R/R_{200}$ for the bulges (red, magenta) and the disks (blue, light blue) of the 469 double-component galaxies within $2.5\,R_{200}$. Best-fitting lines are estimated with Hyper-Fit. Squares represent the median values. Disks become bluer with increasing cluster-centric distance (3.7$\sigma$ level), while bulges show no color-radius relation.}
\label{colorsRadius}
\end{figure*}

\begin{figure*}
\centering
\includegraphics[width=0.45\textwidth]{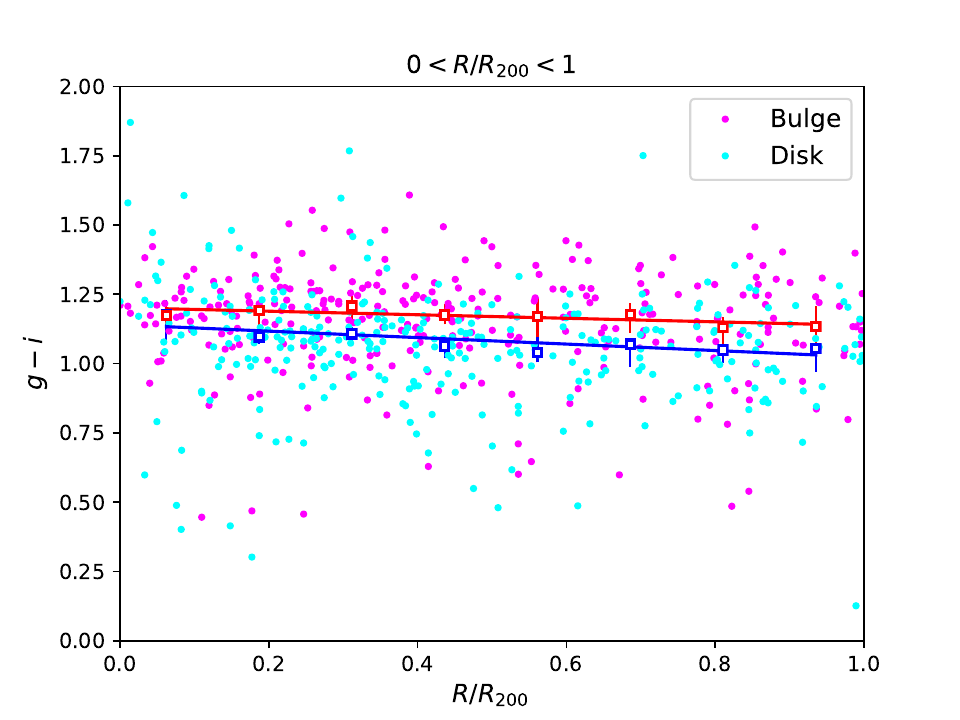}
\includegraphics[width=0.45\textwidth]{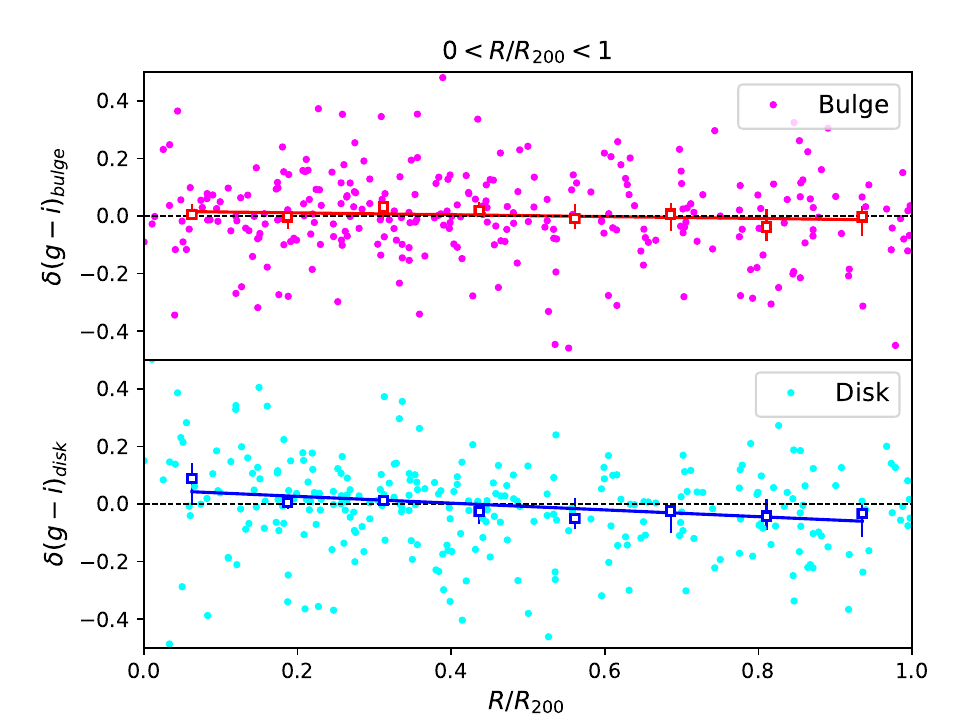}
\includegraphics[width=0.45\textwidth]{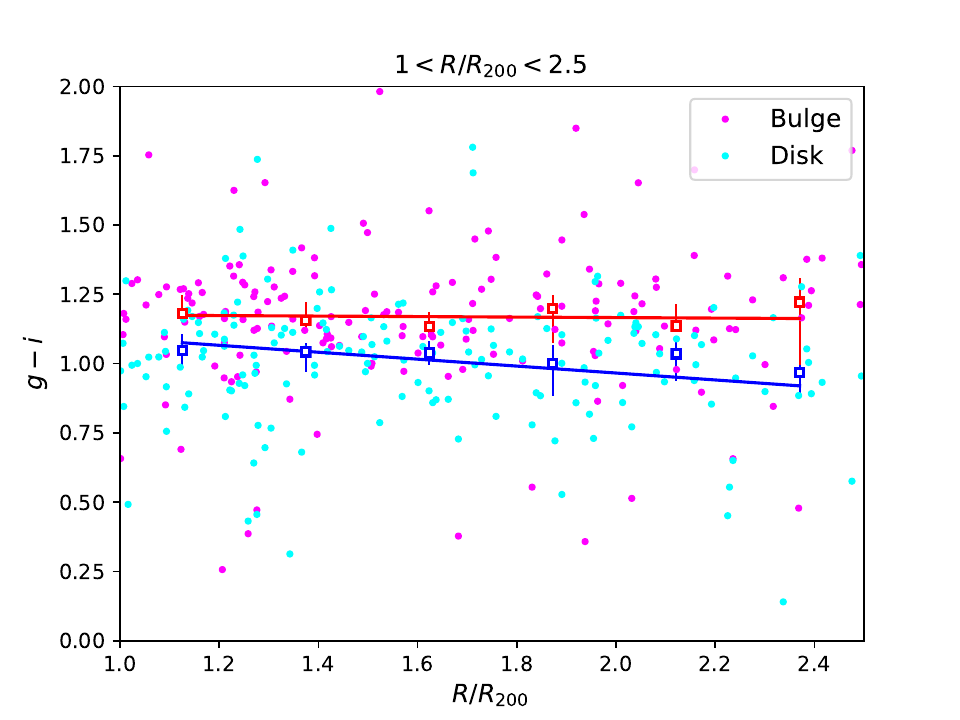}
\includegraphics[width=0.45\textwidth]{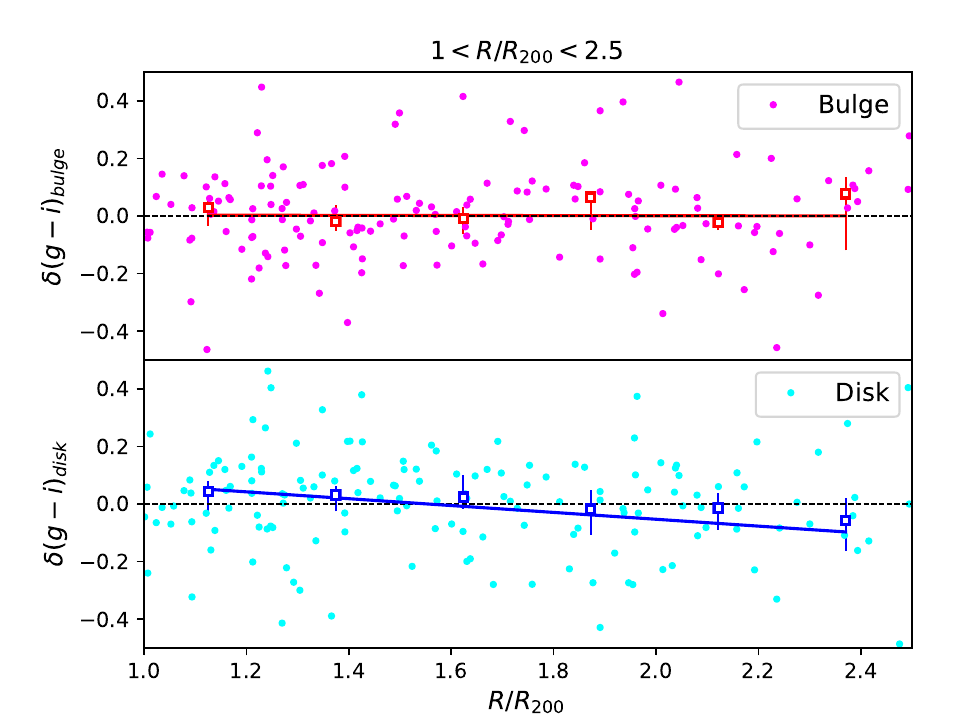}
\caption{$g-i$ colors (\textit{left panels}) and $g-i$ colors decoupled from the color-magnitude relations (\textit{right panels}) versus $R/R_{200}$ for the bulges and the disks at $0\leq R/ R_{200}<1$ (\textit{top panels}) and at $1\leq R/ R_{200}<2.5$ (\textit{bottom panels}). A significant color-radius relation ($\sim 3\sigma$ level) is found for disks within $R_{200}$.}
\label{colorsRadius1}
\end{figure*}

\begin{table*}
 \centering
  \caption{Color-Radius relations as a function of $R/R_{200}$. Column 1 lists the sample, column 2 the radial distance, column 3 the number of galaxies, columns 4/5 the bulge/disk slopes for the colors and columns 6/7 the bulge/disk slopes for the decoupled colors from the color-magnitude relations.} 
  \label{CRslopes}
  \begin{tabular}{@{}lcccccc@{}}
  \hline
Galaxy Sample    & $R/R_{200}$  & $N_{gal}$& $CR_{bulge}$ & $CR_{disk}$ &$\delta CR_{bulge}$ & $\delta CR_{disk}$\\
\hline
Total  & 0.0 $-$ 2.5 & 469 &  $-$0.003$\pm$0.018& $-$0.078$\pm$0.021&$-$0.005$\pm$0.018&$-$0.078$\pm$0.021\\
Virialized  & 0.0 $-$ 1.0&  300& $-$0.064$\pm$0.028&$-$0.117$\pm$0.042&$-$0.032$\pm$0.029 &$-$0.117$\pm$0.040\\
Infalling/Backsplash  & 1.0 $-$ 2.5&169 &  $-$0.009$\pm$0.068&$-$0.125$\pm$0.065&  $-$0.002$\pm$0.066&$-$0.119$\pm$0.064\\
\hline
\end{tabular}
\end{table*}

\begin{table*}
 \centering
  \caption{Spearman correlation test for the decoupled color-radius relations. Column 1 lists the sample, column 2 the radial distance, column 3/4 the Spearman rank correlation coefficients with the associated P-values for bulges, and columns 5/6 the Spearman coefficients and P-values for disks.} 
  \label{CRcorr}
  \begin{tabular}{@{}lccccc@{}}
  \hline
Galaxy Sample     &     $R/R_{200}$  & $(\delta CR)\rho_{bulge}$ &$(\delta CR)P_{bulge}$ & $(\delta CR)\rho_{disk}$ &$(\delta CR)P_{disk}$\\
\hline
Total  & 0.0 $-$ 2.5 & $-$0.02&0.651&$-$0.20&9.92$\times10^{-6}$\\
Virialized  & 0.0 $-$ 1.0&$-$0.07&0.256&$-$0.19&1.23$\times10^{-3}$\\
Infalling/Backsplash  & 1.0 $-$ 2.5& $-$0.04&0.643&$-$0.11&0.16\\
\hline
\end{tabular}
\end{table*} 

\subsection{Trends with cluster radius}
\label{Trends with cluster radius}
In order to understand the formation of S0 galaxies in clusters, we aim to explore the influence of the environment on the colors of the bulges and the disks. To this end, we study whether the bulge and disk colors correlate with the projected cluster-centric distance of the galaxy $R$ for the 469 double-component galaxies within $2.5\,R_{200}$.

The \textit{left panel} of Figure~\ref{colorsRadius} shows the separate bulge and disk $g-i$ colors as a function of $R/R_{200}$. We fit the data with the respective errors using the Hyper-Fit software. The best-fitting line for bulges (in red) has a slope equal to $-$0.003$\pm$0.018, consistent with zero. The slope of the best-fitting line for disks (in blue) is $-$0.078$\pm$0.021, which differs from a zero trend at the $3.7\sigma$ level: disks become redder with decreasing radius. The \textit{right panel} of Figure~\ref{colorsRadius} shows the bulge and disk $g-i$ colors decoupled from the respective color-magnitude relation, $\delta(g-i)_{bulge}$ and $\delta(g-i)_{disk}$, as a function of $R/R_{200}$. The decoupled color $\delta(g-i)$ is given by:
\begin{equation}
\delta(g-i)=(g-i)-(g-i)_{CM}
\end{equation}
where $(g-i)_{CM}$ is the colour difference estimated using the color-magnitude relation from Hyper-Fit. The inner regions of clusters are mainly populated by more luminous and more bulge-dominated galaxies, thus the colors decoupled from the color-magnitude relations offer an unbiased probe to track trends with the cluster radius \citep{Head2014}. Fitting the $\delta(g-i)$ values with Hyper-Fit, the slope of the best-fitting line for the bulges is $-$0.005$\pm$0.018, while for the disks is $-$0.078$\pm$0.021. The trend for bulges is consistently flat, while disk colors become bluer at larger cluster radii. We apply the Spearman correlation test, finding a significant correlation only for disks ($\rho$=$-$0.20, P-value=9.92$\times10^{-6}$). Consistent $\rho$ and P-values are obtained for both the colors and the decoupled colors as a function of $R/R_{200}$.

We study the bulge/disk color-radius relations separately for the virialized galaxy population with $0\leq R/ R_{200}<1$ (\textit{top panels} of Figure~\ref{colorsRadius1}) and the mixed infalling/backsplash galaxy population at $1\leq R/ R_{200}<2.5$ (\textit{bottom panels} of Figure~\ref{colorsRadius1}). Table~\ref{CRslopes} summarizes the slope values of the bulge/disk color-radius relations for the total, the virialized and the mixed infalling/backsplash galaxy samples. The flat trend for bulges is observed for both the virialized and the infalling/backsplash galaxies. The disks show statistically indistinguishable slopes for the virialized and the infalling/backsplash regions. However, both disk color-radius trends are marginal, differing from a zero slope at the $2.8\sigma$ and $\sim2\sigma$ level for the virialized and infalling/backsplash regions, respectively. Table~\ref{CRcorr} reports the results of the Spearman rank correlation test. A statistically significant color-radius correlation is measured only for disks of the virialized galaxies.

In order to explore galaxies primarily in the cluster core and those recently accreted from the field, we divide galaxies into four radial sub-samples: $0\leq R/ R_{200}<0.5$ is a region dominated by the virialized population, $0.5\leq R/ R_{200}<1$ contains a mix of virialized, infalling, and backsplash galaxies, $1\leq R/ R_{200}<2$ is dominated by infalling and backsplash galaxies, while $2\leq R/ R_{200}<2.5$ is dominated by the infalling population. Simulations have shown that these radial regions contain populations of galaxies with different accretion times into the cluster \citep{Gill2005,Mahajan2011,Oman2016}. Table~\ref{colorsBD} lists the bulge/disk median $g-i$ colors for the four galaxy sub-samples. The bulge median colors are consistent at each radial distance, while the disk median colors decrease from the inner to the outer region by 0.096$\pm$0.037 mag.

\begin{figure}
\centering
\includegraphics[width=\columnwidth]{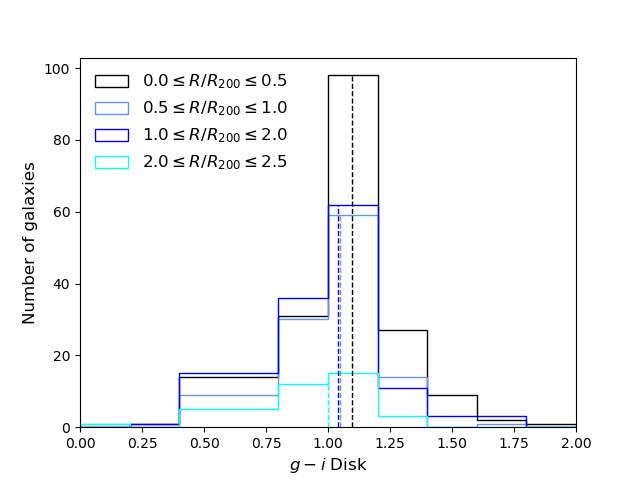}
\caption{Histograms for the disk $g-i$ colors for galaxies at different radial ranges. The dashed lines represent the median $g-i$ colors. The disk colors decrease from the inner to the outer region by 0.096$\pm$0.0037 mag.}
\label{histogramDiskColors}
\end{figure}

Figure~\ref{histogramDiskColors} shows the histograms for the disk colors at the different radial ranges. To assess the statistical significance, we apply the Kolmogorov$-$Smirnov test (K$-$S test; \citealp{Lederman1984}), which tests the null hypothesis that the bulge/disk $g-i$ color distributions of two galaxy sub-samples are drawn from the same parent population. The test returns the probability of measuring the observed difference between two cumulative distributions if they were drawn from the same parent population. P-values smaller than 0.05 indicate that the two distributions are unlikely to be drawn from the same parent distribution. Table~\ref{KSresults} lists the P-values for the comparisons between the four sub-samples. The high P-values for the bulge $g-i$ distributions indicate that they are likely to belong to the same parent population. 

The disk colors for the virialized galaxies inside the cluster core at $R< 0.5\,R_{200}$ are observed to be significantly different compared to the disk colors distributions for galaxies at the larger distances $0.5\leq R/R_{200}<1$, $1\leq R/R_{200}<2$ and $2\leq R/R_{200}<2.5$. We observe no differences for the disk colors between the virialized galaxies outside the cluster core, the infalling/backsplash galaxies and the infalling galaxies. These results imply that the disk color-radius relation for virialized galaxies is primarily driven by galaxies in the cluster core at $0\leq R/ R_{200}<0.5$.

\begin{figure}
\centering
\includegraphics[width=\columnwidth]{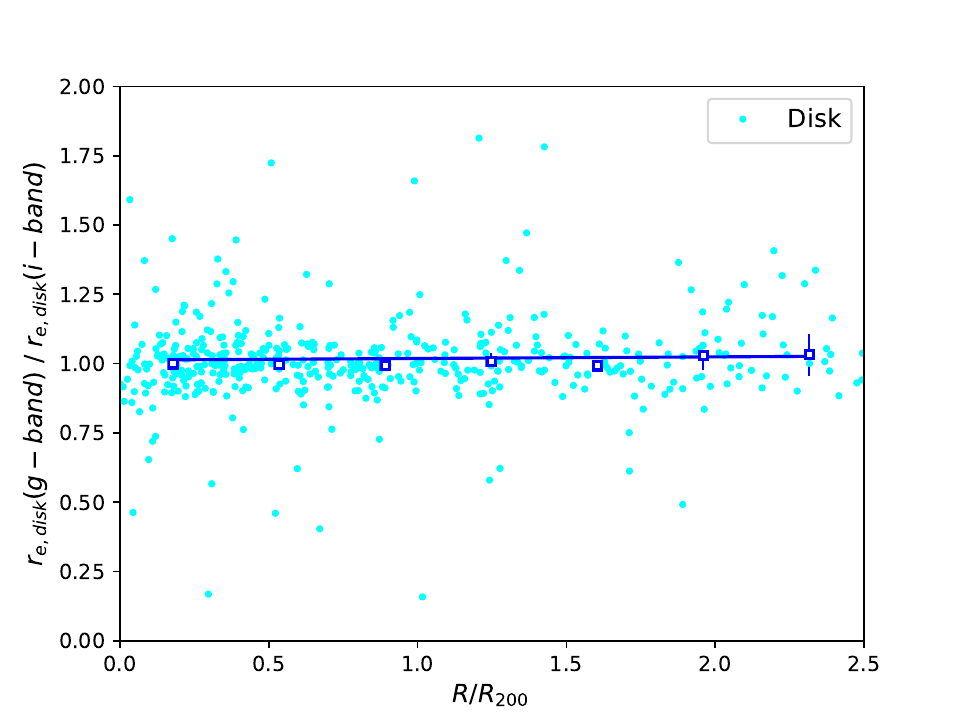}
\caption{$g$-to-$i$-band $r_{e,disk}$ ratio as a function of $R/R_{200}$ for the 469 double-component galaxies within $2.5\,R_{200}$. We find no significant change of $r_{e,disk}$ with the environment.}
\label{ReDiskTrend}
\end{figure}

Finally, to understand which environment mechanisms mainly cause the disk color-radius correlation, we explore whether the disk effective radius changes as a function of $R/R_{200}$. A decrease of the $g$-to-$i$-band $r_{e,disk}$ ratio at smaller cluster-centric distances would suggest that the quenching process is spatially-dependent. Figure~\ref{ReDiskTrend} shows the $g$-to-$i$-band $r_{e,disk}$ ratio as a function of $R/R_{200}$. The slope of the best-fitting line is 0.006$\pm$0.014, consistent with a flat trend. No correlation of the $g$-to-$i$-band $r_{e,disk}$ ratio with the environment is also found according to the Spearman test ($\rho$=0.06, P-value=0.232). The same result is also observed for only the virialized galaxies within $1\,R_{200}$.


\begin{table}
 \centering
  \caption{Comparison of the bulge/disk colors for galaxies at different radial ranges. Column 1 lists the radial range, column 2 the number of galaxies and columns 3/4 the bulge/disk median $g-i$ colors with the standard errors.} 
  \label{colorsBD}
  \begin{tabular}{@{}cccc@{}}
  \\
  \hline
 $R/R_{200}$  & $N_{gal}$& $<g-i>_{bulge}$ & $<g-i>_{disk}$ \\
\hline
 0.0 $-$ 0.5 & 184 &  1.191$\pm$0.009& 1.096$\pm$0.010\\
 0.5 $-$ 1.0 & 116 &  1.151$\pm$0.014& 1.048$\pm$0.013\\
 1.0 $-$ 2.0 & 133 & 1.176$\pm$0.013& 1.038$\pm$0.014\\
 2.0 $-$ 2.5 & 36 &1.203$\pm$0.027 & 1.000$\pm$0.027\\
\hline
\end{tabular}
\end{table}

\begin{table}
 \centering
 \caption{K$-$S test results. Columns 1/2 lists the radial ranges of the compared galaxy samples and columns 3/4 the P-values for the bulge/disk $g-i$ color distributions.} 
  \label{KSresults}
  \begin{tabular}{@{}cccc@{}}
  \hline
 Sample 1 ($R/R_{200}$)  &  Sample 2 ($R/R_{200}$)    &    $P_{bulge}$ & $P_{disk}$  \\
\hline
0.0 $-$ 0.5 & 0.5 $-$ 1.0 & 0.0971 &  0.0057\\
0.0 $-$ 0.5 & 1.0 $-$ 2.0 & 0.8118 &  0.0002\\
0.0 $-$ 0.5 & 2.0 $-$ 2.5 & 0.7000 &  0.0117\\
0.5 $-$ 1.0 & 1.0 $-$ 2.0& 0.6529 &  0.7404\\
0.5 $-$ 1.0 & 2.0 $-$ 2.5 & 0.4565 &  0.1720\\
1.0 $-$ 2.0 & 2.0 $-$ 2.5 & 0.9007 &  0.6373\\
\hline
\end{tabular}
\end{table}

\subsection{Trends with projected phase space: disentangling backsplash and infalling galaxies}
\label{Trends with projected phase space: disentangling backsplash and infalling galaxies}
In Section~\ref{Trends with cluster radius} we find that virialized galaxies within $0.5\,R_{200}$ have a disk $g-i$ color distribution that differs significantly from galaxies at $0.5\leq R/R_{200}<2.5$. We find no differences between galaxy samples binned by radius outside $0.5\,R_{200}$. However, the use of only radial bins is not sufficient to separate backsplash from infalling galaxies. Both these galaxy populations are found at $1\leq R/ R_{200}<2$, but are predicted to have different velocity distributions \citep{Gill2005}. In this Section we aim to disentangle backsplash and infalling galaxies. A significant difference between the disk $g-i$ color distributions of backsplash galaxies and of infalling galaxies suggests a short timescale for the star formation quenching processes acting within $0.5\,R_{200}$, since one passage through the cluster core is enough to change the disk $g-i$ distribution of backsplash galaxies. On the other hand, similar disk $g-i$ color distributions for backsplash and infalling galaxies suggest a long quenching timescale for the core processes.

\begin{figure}[h!]
\centering
\includegraphics[width=\columnwidth]{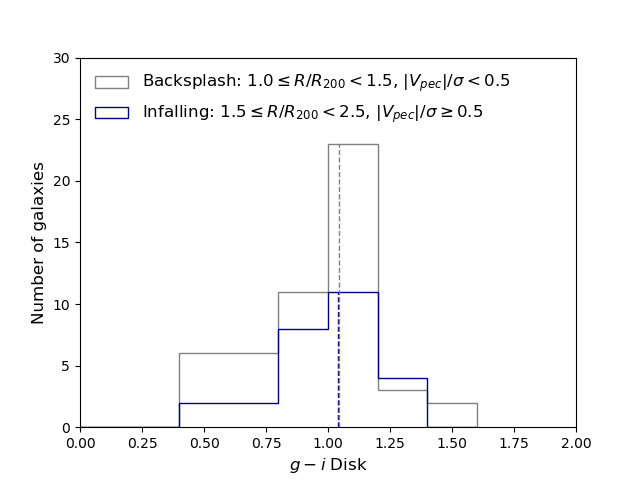}
\caption{Histograms for the disk $g-i$ colors for galaxies in different PPS regions, i,e, backsplash ($1\leq R/ R_{200}<1.5$; $|V_{pec}|/\sigma_{200}<$0.5) and infalling ($1.5\leq R/ R_{200}<2.5$; $|V_{pec}|/\sigma_{200}\geq$0.5). The dashed lines represent similar median $g-i$ colors.}
\label{histogramInfBackColors}
\end{figure}

To separate backsplash from infalling galaxies we use the projected phase space (PPS) diagram, i.e. the galaxy velocity as a function of projected cluster-centric radius, which has been extensively used in the literature (e.g., \citealp{Mahajan2011,Oman2013,Rhee2017,Yoon2017,Owers2019}). Following \citet{Yoon2017}, we select two regions in PPS: a region with $1\leq R/ R_{200}<1.5$ and $|V_{pec}|/\sigma_{200}<$0.5 that is thought to contain a larger proportion of backsplash galaxies, and a region with  $1.5\leq R/ R_{200}<2.5$ and $|V_{pec}|/\sigma_{200}\geq$0.5 that is dominated by infalling galaxies. $V_{pec}$ is the peculiar velocity and $\sigma_{200}$ is the cluster velocity dispersion measured by \citet{Owers2017}. We find 46 backsplash galaxies and 25 infalling galaxies. Figure~\ref{histogramInfBackColors} shows the $g-i$ color histograms for the disks of the backsplash galaxies and the infalling galaxies. Applying the K$-$S test, we find no significant difference between the disk $g-i$ color distributions of the two galaxy populations (P-value = 0.9286). 

\begin{table*}
 \centering
  \caption{Color-Density relations as a function of $R/R_{200}$. Column 1 lists the sample, column 2 the radial distance, column 3 the number of galaxies, columns 4/5 the bulge/disk slopes for the colors and columns 6/7 the bulge/disk slopes for the decoupled colors from the color-magnitude relations.} 
  \label{CDslopes}
  \begin{tabular}{@{}lcccccc@{}}
  \hline
Galaxy Sample    & $R/R_{200}$  & $N_{gal}$& $CD_{bulge}$ & $CD_{disk}$ &$\delta CD_{bulge}$ & $\delta CD_{disk}$\\
\hline

Total  & 0.0 $-$ 2.5 & 469 &  0.036$\pm$0.022& 0.089$\pm$0.026&0.033$\pm$0.021&0.088$\pm$0.026\\
Virialized  & 0.0 $-$ 1.0&  300& 0.047$\pm$0.031&0.118$\pm$0.043&0.041$\pm$0.030 &0.118$\pm$0.043\\
Infalling/Backsplash  & 1.0 $-$ 2.5&169 &  0.087$\pm$0.053&0.018$\pm$0.052&  0.077$\pm$0.052&0.012$\pm$0.052\\
\hline
\end{tabular}
\end{table*}

\begin{table*}
 \centering
  \caption{Spearman correlation test for the decoupled color-density relations. Column 1 lists the sample, column 2 the radial distance, column 3/4 the Spearman rank correlation coefficients with the associated P-values for bulges, and columns 5/6 the Spearman coefficients and P-values for disks.} 
  \label{CDcorr}
  \begin{tabular}{@{}lccccc@{}}
  \hline
Galaxy Sample     &     $R/R_{200}$  & $(\delta CD)\rho_{bulge}$ &$(\delta CD)P_{bulge}$ & $(\delta CD)\rho_{disk}$ &$(\delta CD)P_{disk}$\\
\hline
Total  & 0.0 $-$ 2.5 & 0.05&0.323&\hspace{3mm}0.167&0.0003\\
Virialized  & 0.0 $-$ 1.0&0.03&0.557&\hspace{3mm}0.144&0.0126\\
Infalling/Backsplash  & 1.0 $-$ 2.5& 0.09&0.247&$-$0.004&0.9588\\
\hline
\end{tabular}
\end{table*}

\subsection{Trends with local galaxy density}
\label{Trends with galaxy density} 
With the aim of understanding the relative importance of pre-processing and cluster-specific environmental processes on S0 production, we investigate trends with local galaxy density both inside and outside $R_{200}$. The local galaxy density is measured as the fifth nearest neighbour surface density $\Sigma_{5}$, following a procedure similar to \citet{Brough2013}. 

\begin{figure*}
\centering
\includegraphics[width=0.45\textwidth]{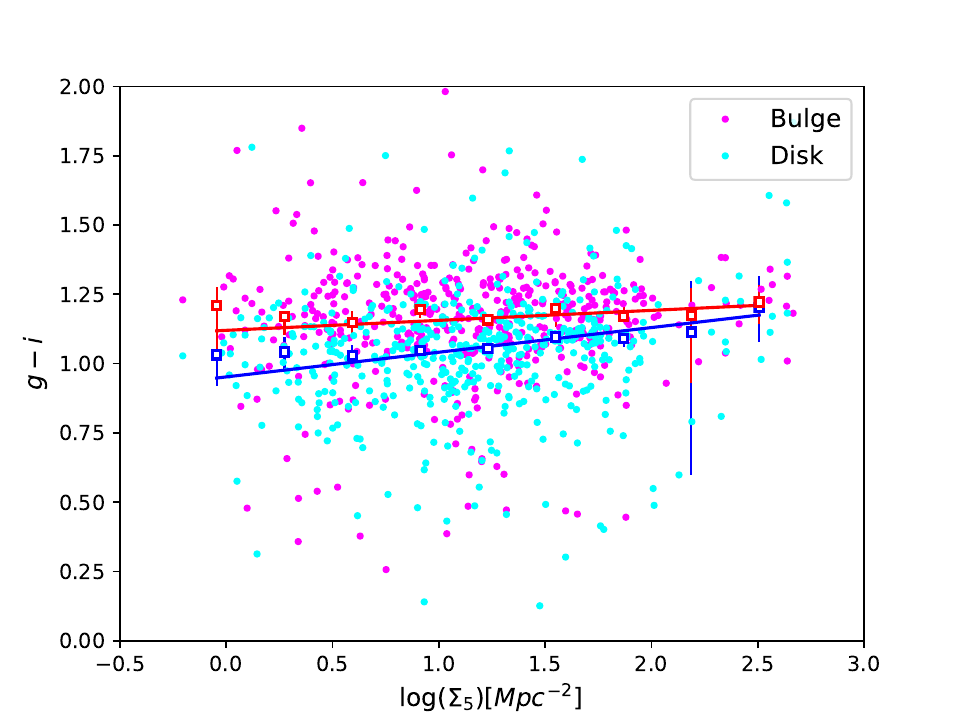}
\includegraphics[width=0.45\textwidth]{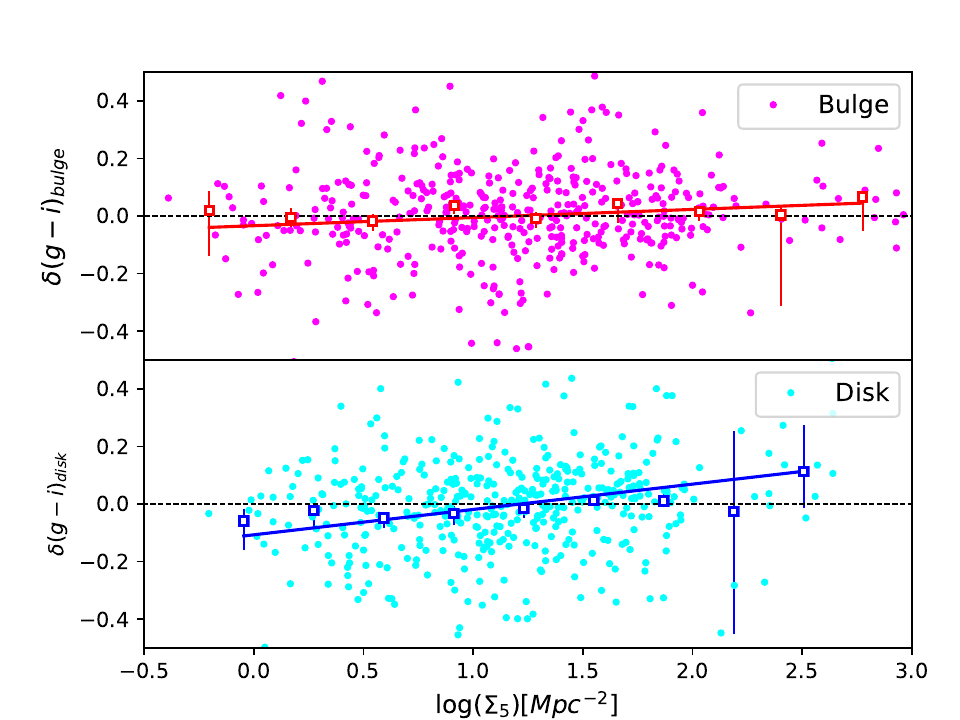}
\includegraphics[width=0.45\textwidth]{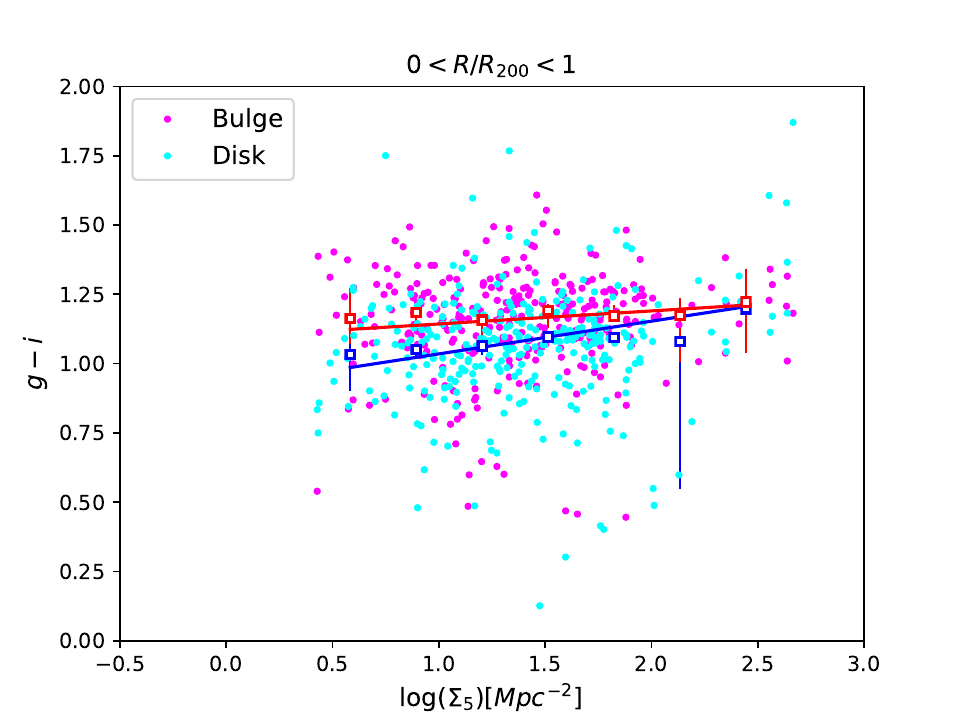}
\includegraphics[width=0.45\textwidth]{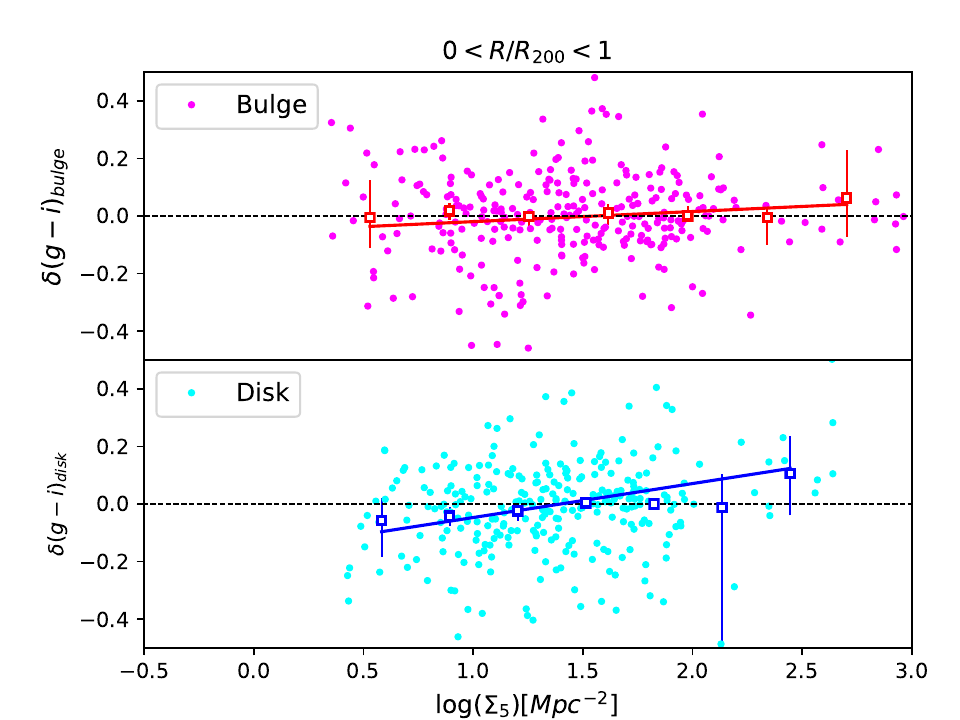}
\includegraphics[width=0.45\textwidth]{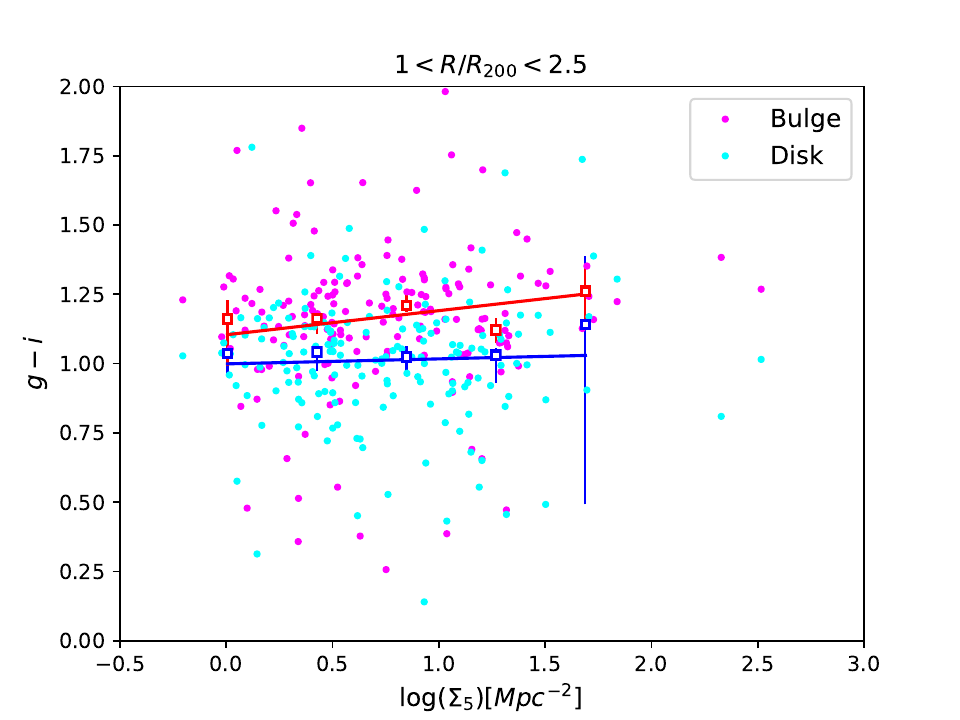}
\includegraphics[width=0.45\textwidth]{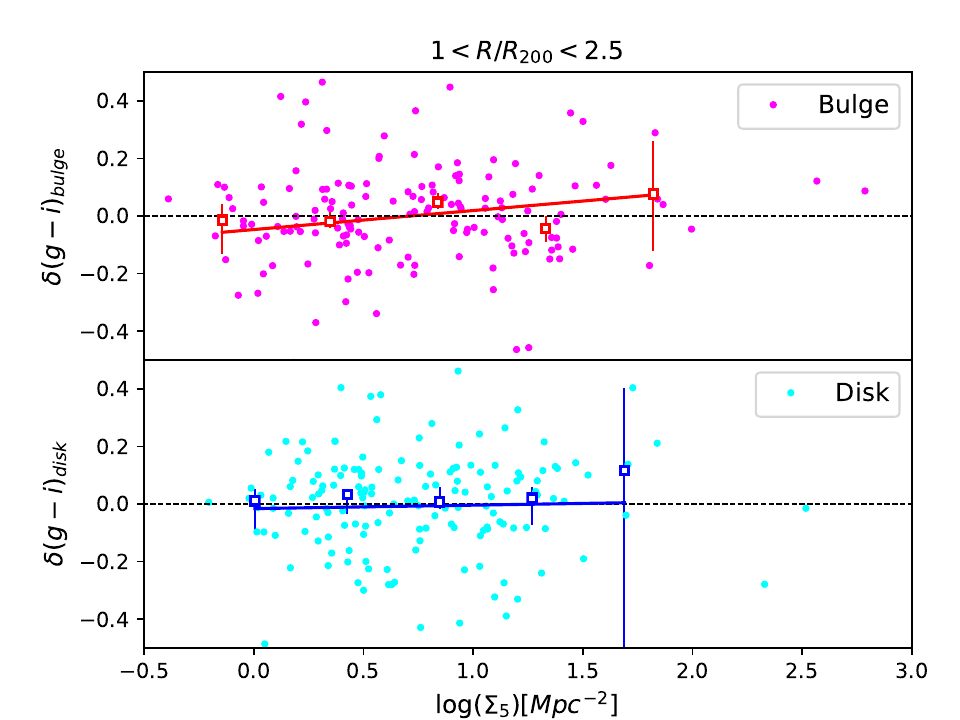}
\caption{$g-i$ colors (\textit{left panels}) and $g-i$ colors decoupled from the color-magnitude relations (\textit{right panels}) versus $\Sigma_{5}$ for the bulges and the disks at  $0\leq R/ R_{200}<2.5$ (\textit{top panels}), $0\leq R/ R_{200}<1$ (\textit{middle panels}) and at $1\leq R/ R_{200}<2.5$ (\textit{bottom panels}). Bulges have a flat color-density relation. Disks become redder with increasing local galaxy density (3.4$\sigma$ level). This trend is primarily due to the disk color-density relation for the virialized galaxies. Beyond $R_{200}$, the disk color-density relation is not significant.}
\label{colorsDensity}
\end{figure*}

Figure~\ref{colorsDensity} shows the bulge/disk $g-i$ colors as a function of the local galaxy density for the total double-component galaxy sample (\textit{top panels}), the virialized galaxies within $R_{200}$ (\textit{middle panels}) and the infalling/backsplash galaxies at $1\leq R/ R_{200}<2.5$ (\textit{bottom panels}). The bulge/disk $g-i$ colors decoupled from the respective color-magnitude relations are shown in the \textit{right panels} of Figure~\ref{colorsDensity}. The slope values of the bulge/disk color-density relations are reported in Table~\ref{CDslopes}. The bulge colors show a flat trend as a function of the local galaxy density for each radial distance. For the total galaxy sample the disk colors become redder with increasing local galaxy density, having a color-density slope that differs from zero at the $3.4\sigma$ level. This trend is primarily due to the disk color-density relation of the virialized galaxies. The disk color-density relation for the infalling/backsplash galaxies is statistically consistent with zero. Table~\ref{CDcorr} lists the results of the Spearman rank correlation test, highlighting no disk color-density correlation for the infalling/backsplash galaxies. These results suggest that the bulge colors are not affected by the environment, while the disk colors become redder with increasing local galaxy density within $R_{200}$. 

In agreement with the finding in Section~\ref{Trends with cluster radius}, we find no significant variation of the $g$-to-$i$-band $r_{e,disk}$ ratio with local galaxy density, measuring the slope of the best-fitting line equal to -0.012$\pm$0.017. However, according to the Spearman test, a marginal correlation is found between the $g$-to-$i$-band $r_{e,disk}$ ratio and local galaxy density ($\rho$=$-$0.09, P-value=0.049).

Finally, consistent results for the color-density relations are obtained if we change the Bayes Factor criterion from $\ln(\rm{BF})>60$ to $\ln(\rm{BF})>5$ for galaxy characterization in Section~\ref{sec:Model selection}. 


\subsection{Comparison with single-component disk-dominated galaxies}
\label{Comparison with disk dominated galaxies}
For double-component galaxies we find that the bulge colors do not correlate with the environment, while the disk colors correlate with the cluster radius and the local galaxy density. However, no significant disk color-density correlation is detected for double-component galaxies beyond $1\,R_{200}$. We aim to understand whether the environment, and in particular pre-processing taking place in the cluster outskirts, affects single-component disk-dominated galaxies. We study the dependence of the $g-i$ colors of the disk-dominated galaxies on the local galaxy density. In Section~\ref{sec:Model selection} we identify 394 single-component galaxies with S\'ersic index $n<2$, i.e. disk-dominated. 

\begin{figure}
\centering
\includegraphics[width=\columnwidth]{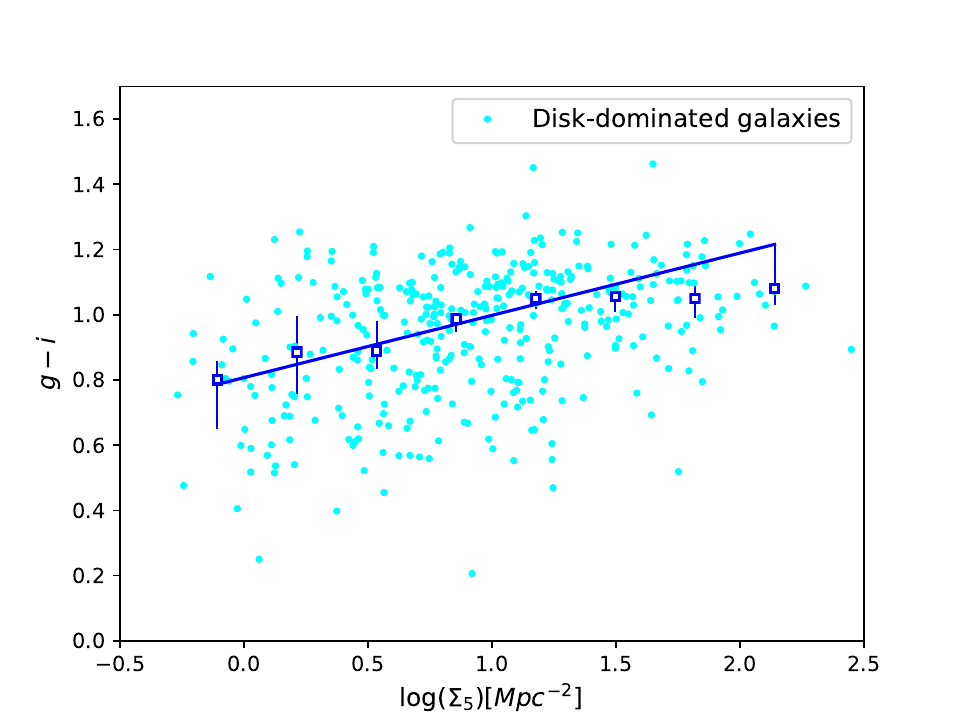}
\includegraphics[width=\columnwidth]{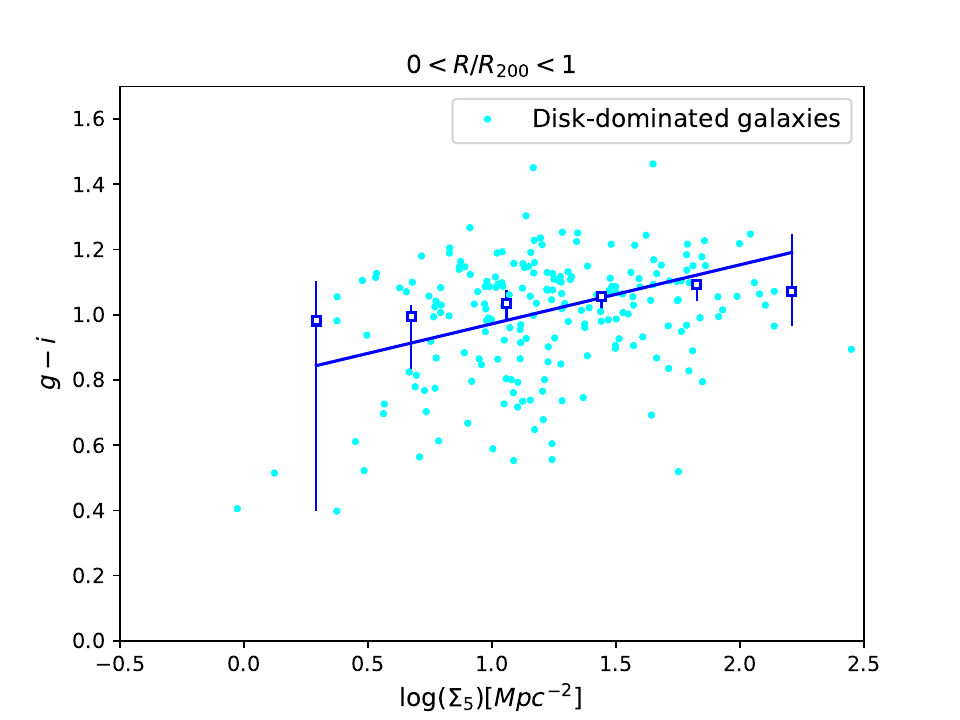}
\includegraphics[width=\columnwidth]{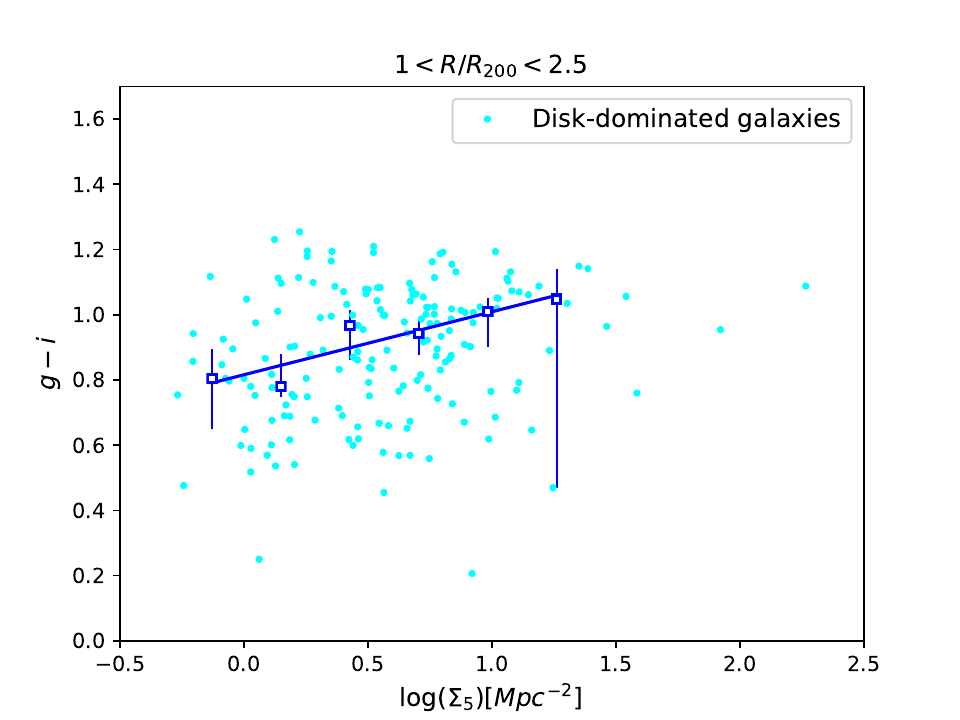}
\caption{$g-i$ colors versus $\Sigma_{5}$ for single-component disk-dominated galaxies at $0\leq R/ R_{200}<2.5$ (\textit{top panel}), $0\leq R/ R_{200}<1$ (\textit{middle panel}) and at $1\leq R/ R_{200}<2.5$ (\textit{bottom panel}). Disks become redder with increasing local galaxy density (3.3$\sigma$ level). Significant trends are observed for galaxies within and beyond $R_{200}$.}
\label{colorsDensityDiskDominated}
\end{figure}

Figure~\ref{colorsDensityDiskDominated} shows that the colors of the disk-dominated galaxies correlate strongly with the local galaxy density for the total, the virialized and the infalling/backsplash samples. The colors become redder with increasing local galaxy density. Table~\ref{CDslopesDiskDominated} lists the slopes and the results of the Spearman test for the three samples. Since we consider disk-dominated galaxies, there is no need to decouple the colors from their color-magnitude relation. Significant color-density relations at the 3.8$\sigma$ and 3.3$\sigma$ levels are observed for disk-dominated galaxies within and beyond $R_{200}$, respectively.

To understand the discrepancy between the disk color-density correlation for single-component disk-dominated galaxies in the cluster outskirts and no trend for double-component galaxies, we compare the properties of these two galaxy samples beyond $R_{200}$. We observe that disk-dominated galaxies are characterized by lower galaxy stellar masses than the double-component galaxies. The median $\log(M_{*}/M_{\odot})$ difference is 0.614$\pm$0.067. The disks of the disk-dominated galaxies are bluer than the disks of the double-component galaxies, showing a median $g-i$ color difference of 0.116$\pm$0.028 mag.

\begin{table*}
 \centering
  \caption{Color-Density relations as a function of $R/R_{200}$ for single-component disk-dominated galaxies. Column 1 lists the sample, column 2 the radial distance, column 3 the number of galaxies, column 4 the disk-dominated slopes for the colors and columns 5/6 the Spearman rank correlation coefficients with the associated P-values.} 
  \label{CDslopesDiskDominated}
  \begin{tabular}{@{}lccccc@{}}
  \hline
Galaxy Sample    & $R/R_{200}$  & $N_{gal}$& $CD_{disk}$ & $(CD)\rho_{disk}$ &$(CD)P_{disk}$ \\
\hline

Total  & 0.0 $-$ 2.5 & 394 &  0.191$\pm$0.058&0.33&1.36$\times10^{-11}$\\
Virialized  & 0.0 $-$ 1.0& 214 & 0.181$\pm$0.049&0.22&1.24$\times10^{-3}$\\
Infalling/Backsplash  & 1.0 $-$ 2.5& 180 &  0.193$\pm$0.059&0.23&1.77$\times10^{-3}$\\
\hline
\end{tabular}
\end{table*}

\section{Discussion}
In this paper, we seek to understand the role of environment on the formation of S0 galaxies. We focus specifically on galaxy clusters and their poorly studied outskirts, where we investigate separately the colors of bulges and disks as proxies for their stellar population properties. In this Section, we discuss our results and how they can be interpreted in the context of cluster-specific environmental processes, as well as pre-processing in groups prior to accretion.

We use the 2D photometric bulge-disk decomposition to characterize the double-component galaxies out to $2.5\,R_{200}$ in Section~\ref{sec:Model selection}. We are able to measure reliably the bulge and disk colors for 469 galaxies out of the 1083 double-component galaxies with $\ln(BF)>60$. Most (404) of the excluded galaxies have unresolved bulge and disk radial parameters. These galaxies are less massive (by 0.145$\pm$0.037 dex in $\log(M_{*}/M_{\odot})$) and smaller in size (by 0.469$\pm$0.091 arcsec in single-component $r_{e}$) compared to the reliable double-component sample. Thus, this exclusion limits our results to massive and large galaxies.

In Section~\ref{Color-magnitude relations} we observe that $67\pm2$\% of the bulges have redder colors than the disks with a median $g-i$ offset of 0.11$\pm$0.02 mag. This result is in agreement with previous studies on the cluster environment \citep{Hudson2010,Head2014}. In particular, \citet{Head2014} observed a similar median $g-i$ offset of 0.097$\pm$0.004 mag for bulges being redder than the disks. The remaining $33\pm2$\% of the bulges are found to be bluer than the disks. This is observed for both virialized and backsplash/infalling galaxies. The bluer color of the bulge relative to the disk may be caused by either a younger or more metal-poor bulge stellar population, or a combination of both. However, the study of colors alone is not able to distinguish these scenarios. On the other hand, spectroscopic observations, which can break the age-metallicity degeneracy, have found evidence for bulges containing stellar populations that are younger than their surrounding disks \citep{Johnston2014,FraserMcKelvie2018}. For 13 S0 galaxies in the Virgo cluster, \citet{Johnston2014} observed bulges to be younger and more metal-rich than the disks. \citet{FraserMcKelvie2018} studied 279 S0 field galaxies, finding a galaxy population with bulges younger than the disks, mainly for low mass galaxies. The blue, younger bulges may be the result of ram-pressure stripping, or galaxy-galaxy interactions. Ram-pressure stripping removes gas from the disk, quenching the star formation in the disk while it can continue for longer in the bulge. This leads to an older, redder disk compared to the younger, bluer bulge \citep{Fujita1999}. Galaxy-galaxy interactions that trigger star formation activity in the galaxy central region are able to lead to a bulge younger than the disk \citep{Bekki2011}.

A statistically significant (4.6$\sigma$ level) color-magnitude relation for bulges is observed within $2.5\,R_{200}$. According to the Spearman test, this correlation for bulges is found within and beyond $1\,R_{200}$. Disks show no color-magnitude correlation for both the inner and the outer cluster regions. These results suggest that the bulge component is mainly responsible for the global color-magnitude trend for cluster S0 galaxies.

Within $1\,R_{200}$ we measure color-magnitude slopes of $-$0.056$\pm$0.013 for bulges and 0.003$\pm$0.017 for disks, observing a significant color-magnitude relation for bulges (4.3$\sigma$ level) and a flat trend for disks. Both \citet{Hudson2010} and \citet{Head2014} found a steeper color-magnitude relation for bulges compared to the disks within $\sim 1\,R_{200}$ (see their Figure 11 and Figure 6, respectively). \citet{Hudson2010}, in agreement with our results, found a flat trend for disks, implying that the global color-magnitude relation for cluster galaxies is only due to the change in bulge color. On the other hand, \citet{Head2014} measured significant $g-i$ color-magnitude slopes of $-$0.060$\pm$0.004 and of $-$0.036$\pm$0.005 for bulges and for disks, respectively. In this latter scenario both components contribute to the global color-magnitude relation. \citet{Head2014} explored that this difference compared to \citet{Hudson2010} is not a consequence of selection bias, possibly making the Coma cluster a particular case where the disk component affects the global trend also at $R<1\,R_{200}$. The color-magnitude slopes measured by \citet{Head2014} are consistent with our results within $<2\sigma$. However, we do not find a significant color-magnitude relation for disks, which might be due to the quality of our data. Comparing Figure~\ref{colorsCMRradial} with Figure 6 of \citet{Head2014}, they observed much lower scatter for the two components and probed the Coma cluster about $1$ mag deeper.

The analyses of the bulge/disk colors as a function of the environment metrics in Sections~\ref{Trends with cluster radius} and~\ref{Trends with galaxy density} reveal that the bulge colors do not correlate with environment metrics, while the disk colors become redder with decreasing cluster-centric radius and increasing local galaxy density. The fact that the bulge component does not correlate with the environment, suggests that galaxy interactions and mergers are not primarily responsible for the formation of S0 galaxies in clusters, since these mechanisms can cause star formation bursts in bulges \citep{Bekki2011}. 

A correlation between the disk colors and the cluster-centric radius is measured for galaxies within $1\,R_{200}$. Both \citet{Hudson2010} and \citet{Head2014} found a steeper color-radius relation for disks compared to the bulges for the inner cluster region (see their Figure 15 and Figure 10, respectively). For the decoupled color-radius relations \citet{Head2014} observed slopes of $-$0.029$\pm$0.012 for bulges and $-$0.046$\pm$0.013 for disks, consistent within $1\sigma$ with our measured values of $-$0.032$\pm$0.029 for bulges and $-$0.117$\pm$0.040 for disks. We observe the disk $g-i$ colors to become bluer by 0.05 mag from the cluster core to the virial radius. This value is consistent with the results of \citet{Hudson2010} and \citet{Head2014} who observed a decrease of $\sim$ 0.10 mag in $B-R$ and $\sim$ 0.04 mag in $g-i$, respectively. Applying the $K-S$ test, we observe that the disk colors within the cluster core at $R<0.5\,R_{200}$ are significantly different with respect to the disk color distributions for the galaxies outside the cluster core at $0.5\leq R/ R_{200}<1$, for the infalling/backsplash galaxies at $1\leq R/ R_{200}<2$ and for the mainly infalling galaxies at $2\leq R/ R_{200}<2.5$. \citet{Hudson2010} found a similar result for the disk colors, observing a clear difference between the cluster core at $R<0.3\,R_{200}$ and the outer region at $0.3<R/R_{200}<1$. For the clusters analyzed in this work, \citet{Owers2019} found evidence of star formation quenching induced by ram-pressure stripping acting on infalling spiral galaxies within $0.5\,R_{200}$. Most of these galaxies were observed to be partially stripped. All these results indicate that physical processes acting within the cluster core are important for transforming the star-forming properties of galaxies. 

In Section~\ref{Trends with galaxy density} we also observe a strong disk color-density trend for the virialized galaxies within $R_{200}$. Oppositely to these results, \citet{Lackner2013} measured no correlation between the disk colors and the local galaxy density for rich clusters. However, they found a weak correlation between disk mass and the cluster crossing time as environment metric, suggesting that morphological transformation takes places in rich clusters.

To  understand  which  environment mechanisms mainly drive the disk color-radius/density correlations, we investigate the $g$-to-$i$-band $r_{e,disk}$ ratio as a function of cluster-centric distance and local galaxy density. We observe flat trends, measuring a marginal correlation only for the local galaxy density according to the Spearman test. This might indicate that the physical process that is driving the disk color gradients with environment metrics does not cause spatially-dependent quenching in the disks, suggesting strangulation to be mainly responsible. On the other hand, the quenching process might affect the disk structure, like ram-pressure stripping, but the photometric measurements are not sensitive enough and/or the galaxy sample is not large enough to measure a statistically significant correlation between the $g$-to-$i$-band $r_{e,disk}$ ratio and the environment metrics. In this context, deeper ultraviolet and $u$-band images as well as integral field spectroscopic observations would be able to disentangle the predicted signatures of the various environmental mechanisms. 

The use of the projected phase space instead of radial-only bins in Section~\ref{Trends with projected phase space: disentangling backsplash and infalling galaxies} helps us to discern between backsplash galaxies and infalling galaxies. We find no significant difference between the disk color distributions of these two galaxy populations, suggesting a long timescale for star formation quenching in the disks of S0 galaxies. This finding is in agreement with a scenario where a pure backsplash model cannot be the only responsible driver for galaxy transformations in clusters \citep{Rines2005,Pimbblet2006}. Slow mechanisms, such as strangulation, or fast core-related mechanisms but acting on multiple passages, such as ram-pressure stripping causing partial gas removal in the disk at each passage, can be responsible for the star formation quenching on a long timescale. We identify 46 backsplash and 25 infalling galaxies, thus larger galaxy samples are needed to confirm this result and they will be possible in the context of the Wide Area VISTA Extra-galactic Survey (WAVES; \citealp{Driver2016}). Moreover, the selection of backsplash galaxies using the PPS diagram is still limited by the possible contamination of infalling galaxies in the PPS backsplash region \citep{Oman2013,Rhee2017}. The information from the PPS combined with measurements of H I gas in galaxies offers a more powerful tool to identify true backsplash galaxies \citep{Jaffe2015,Yoon2017}. 

Within and beyond $R_{200}$ the disk colors show a statistically indistinguishable trends as a function of cluster-centric distance. Both the slopes of the disk color-radius relations are marginally different from zero. According to the Spearman rank test a significant correlation is found within $R_{200}$, however, no significant correlation is measured for galaxies at $1\leq R/ R_{200}<2.5$. Given the relatively small measured slope, a larger sample of galaxies is required to reliably determine if the color-radius relations in the virialized and infalling/backsplash regions are significantly different from a flat trend. A larger galaxy sample is also needed to understand whether the disk color-radius relation in the virialized region is stronger compared to that for galaxies beyond $R_{200}$.

No correlation is observed for the disk colors as a function of local galaxy density beyond $R_{200}$. This flat disk color-density trend might suggest that double-component galaxies are not affected by pre-processing in the cluster outskirts. It is also possible that pre-processing occurred a long time ago in groups and it is not visible in the disk color-radius/density relations. Our finding of no disk color-density correlation for infalling/backsplash double-component galaxies in the cluster outskirts disagrees with the result of \citet{Lackner2013}, who observed a significant disk color-density trend for galaxies in poor groups. A larger sample of galaxies combined with deeper images for the 2D photometric bulge-disk decomposition and the study of ultraviolet colors might help us to further investigate pre-processing in clusters. 

In Section~\ref{Comparison with disk dominated galaxies} for single-component disk-dominated galaxies beyond $R_{200}$ we detect a disk color-density relationship whereby disks become redder with increasing local density. Taking the color of these disk-dominated systems as a proxy for their star forming activity, we interpret this color-density relation as being caused by the quenching of star formation in galaxies in higher density environments. A consistent conclusion was drawn by \citet{Lewis2002}, who used EW(H$\alpha$) as a more direct probe of star formation rate of galaxies in the outskirts of seventeen 2dFGRS clusters. \citet{Lewis2002} found star formation quenching for galaxies in the outer cluster regions tracing a decrease of star formation rate with increasing local galaxy density. The single-component disk-dominated galaxies in this study are, on average, both bluer (by 0.116$\pm$0.028 mag in $g-i$ color) and less massive (by 0.614$\pm$0.067 dex in stellar mass) than those in the double-component sample. These lower mass galaxies are likely to be more susceptible to environment-related quenching than their more massive two-component counterparts, and this may lead to a stronger color-density relation. 

The detection of a color-density relation suggests that single-component disk-dominated galaxies are affected by pre-processing, where more subtle environmental processes are at play. Evidence of pre-processing is observed for the disks of infalling spiral galaxies in galaxy clusters at
intermediate-redshift $0.5\leq z\leq0.8$ \citep{Cantale2016}. Effects of star formation quenching are observed in groups \citep{Rasmussen2012,Schaefer2017,Barsanti2018,Schaefer2019}, where
mild ram-pressure stripping \citep{Hester2006} and mergers \citep{Taranu2013} can act to quench star-forming galaxies, leading to the observed color-density relation in the cluster outskirts.

The fact that we only observe a significant disk color-density relation within $R_{200}$ for the two-component galaxies, while we do observe a disk color-density relation in the cluster outskirts for single-component disk-dominated galaxies, indicates that multiple environmental processes may affect disks. We reach a similar conclusion to that of \citet{Lackner2013}; pre-processing in the cluster outskirts quenches star formation in single-component disk-dominated galaxies, but the formation of S0 galaxies is mainly driven by processes in the cluster core. This conclusion is supported by the results of previous studies, where it has been suggested that star formation quenching and morphological transformation are separate mechanisms and act on different timescales \citep{Lewis2002,McIntosh2004,vandenBosch2008,Bamford2009,Skibba2009}.

Overall, our results agree with the outcomes from the simulations for bulges and disks in rich clusters of \citet{Taranu2014}. Their model predictions are compared with bulge and disk observations of colors and Balmer line indexes for cluster galaxies. They found that best fitting models agree with the bulge component not depending on the environment and the disk colors becoming bluer with increasing cluster-centric distance. Models where quenching occurs only within $0.5\,R_{200}$ yield better fits for the observed disk colors. Consistently, we detect no change of bulge colors with environment metrics, while the correlation for disk colors is mainly driven by galaxies in the cluster core within $0.5\,R_{200}$. \citet{Taranu2014} observed that models with short timescales $\leq 1$ Gyr for star formation quenching do not agree with the observations, producing steep changes for the cluster-centric radial gradients in the disk colors and Balmer line indexes. Instead, long quenching timescales are preferable, suggesting strangulation as the main driver of star formation quenching in galaxy disks compared to a rapid ram-pressure stripping causing the total gas removal at the first passage. We find similar disk color distributions for backsplash and infalling galaxies, suggesting a long quenching timescale. Mechanisms such as strangulation or multiple ram-pressure strippings causing partial gas removal are consistent with our results. Finally, \citet{Taranu2014} found that better fits are obtained if models with pre-processing acting on galaxies in low mass groups prior infalling in the rich clusters are implemented. We observe evidence of pre-processing for single-component disk-dominated galaxies.

\section{Summary and conclusions}
\label{Summary and conclusions}
Our aim is to understand how environmental processes and pre-processing act on the formation of S0 galaxies in clusters. We study eight low-redshift massive clusters, with deep and complete spectroscopy from \citet{Owers2017} that covers both the virialized and infalling regions. We explore the colors of the bulges and disk separately, specifically the color-magnitude relations and the dependence on environment metrics, such as the projected cluster-centric distance and the local galaxy density within $2.5\,R_{200}$.  

The selection of the double-component galaxy sample is based on the 2D photometric bulge-disk decomposition and on a Bayesian approach. We use the packages {\sc ProFound} for source detection and {\sc ProFit} for galaxy light modelling on SDSS and VST/ATLAS images. We fit 1795 cluster galaxies, finding 1655 galaxies having both single-/double-component fits converging in the $r$-band. We select double-component galaxies with Bayes Factor $\ln(BF)>60$, combined with flags on the $r_{e,bulge}$, $r_{e,disk}$, bulge-to-total flux ratio and $n_{bulge}$ values to exclude unphysical and unreliable solutions. We find 469 reliable double-component galaxies, 404 double-component galaxies with unresolved radial parameters and 782 single-component galaxies. Galaxies with unresolved radial parameters are characterized by lower galaxy stellar mass and size compared to reliable double-component galaxies, limiting our results to double-component galaxies with a large mass and size. In future studies a deeper and with higher quality imaging is needed to explore the decomposition of low mass galaxies. Reliable double-component galaxies represent a consistent sample of visually selected S0 galaxies. Single-component galaxies are divided into 388 bulge-dominated with $n>2$ and 394 disk-dominated galaxies with $n<2$.

In the context of cluster S0 galaxies our main results are:
\begin{itemize}
    \item 67$\pm$2\% of bulges are found to be redder than their surrounding disks with a median offset of 0.11$\pm$0.02 mag in $g-i$. The remaining 33$\pm$2\% of the bulges are bluer than the disks. 
    \item Only the bulge component shows a $g-i$ color-magnitude correlation for galaxies at $0\leq R/ R_{200}<2.5$ significant at the 4.6$\sigma$ level and according to the Spearman rank correlation test. Disks show a flat color-magnitude relation. The same trends for bulges and disks are observed within the inner cluster region at $0\leq R/ R_{200}<1$ and for the outer galaxy sample at $1\leq R/ R_{200}<2.5 $.
    \item Disk $g-i$ colors become bluer with increasing projected cluster-centric distance (3.7$\sigma$ level), while the color-radius relation is flat for bulges. The slopes of the virialized and the infalling/backsplash regions are statistically indistinguishable, and both are only marginally different from a zero slope. However, the Spearman rank test indicates a significant disk color-radius correlation within $R_{200}$.
    \item The disk colors decrease from the inner cluster region to the outskirts by 0.096$\pm$0.037 mag. According to the K$-$S test, the $g-i$ color distributions of bulges do not show a significant difference as a function of $R/R_{200}$. On the other hand, the disk colors for the virialized galaxies inside the cluster core at $R< 0.5\,R_{200}$ are observed to be significantly different ($P_{disk}$-values $<0.05$) when compared to the disk colors distributions for the galaxies outside the core at $0.5\leq R/ R_{200}<1$, for the infalling/backsplash galaxies at $1\leq R/ R_{200}<2$ and for the infalling galaxies at $2\leq R/ R_{200}<2.5$.
    \item The $g$-to-$i$-band $r_{e,disk}$ ratio does not change with cluster-centric distance and local galaxy density. 
    \item Analyzing the projected phase space, we do not find a significant difference between the disk color distributions of backsplash galaxies with $1\leq R/ R_{200}<1.5$ and $|V_{pec}|/\sigma_{200}<$0.5, and infalling galaxies at $1.5\leq R/ R_{200}<2.5$ and $|V_{pec}|/\sigma_{200}\geq$0.5. 
    \item Disks become redder with increasing local galaxy density (3.4$\sigma$ level), while the color-density relation is flat for bulges. The change in color for disks is primarily due to the disk color-density relation for the virialized galaxies within $R_{200}$. For the infalling/backsplash galaxies beyond $R_{200}$, the disk color-density relation is not significant.
    \item For single-component disk-dominated galaxies, we detect a significant color-density correlation within and beyond $R_{200}$. At large cluster distances, disk-dominated galaxies are observed to be less massive and with bluer disks than double-component galaxies. 
\end{itemize}

In conclusion, using a complete spectroscopic sample of cluster galaxies that probes the cluster core and outskirts, we are able to disentangle the environmental processes acting on the bulges and the disks in the different cluster regions. Due to the nature of the analyzed galaxy sample, this work is only able to inform specifically on the formation scenarios of S0 galaxies in clusters. For the bulges we do not detect any environment correlation. For the disks we find that core processes within $0.5\,R_{200}$ play an important role in the star formation quenching of double-component galaxies, in agreement with previous studies. Our results indicate that processes that act on longer timescales, such as strangulation or multiple ram-pressure strippings, are the main drivers. We observe no evidence of pre-processing acting on the disk colors of double-component galaxies, indicating that this effect is small or happened a long time ago compared to core processes. We assess that processes occurring in the cores of clusters are primarily responsible for the formation of S0 galaxies compared to pre-processing. On the other hand, evidence of pre-processing is observed for single-component disk-dominated galaxies. These latter galaxies are less massive and have bluer disks compared to the double-component galaxies beyond $R_{200}$. Overall, our results suggest a scenario where pre-processing starts to stop star formation in low mass single-component disk-dominated galaxies, as soon as they are accreted into the cluster outskirts. The formation of more massive double-component galaxies instead is driven by core processes that are effective in the cluster central region.

The study of the colors alone is not sufficient to inform about the ages and metallicities of the galaxy components. In this context, the galaxy characterization based on the 2D bulge-disk decomposition performed in Section~\ref{2D bulge-disk decomposition} has been used in studies of the SAMI cluster galaxies \citep{Oh2020,Barsanti2020}. The combination of the photometric and spectroscopic information allows us to analyze the stellar population properties of the bulges and the disks, including their ages and metallicities. New simple and rapid to run galaxy spectral energy distribution packages, such as {\sc ProSpect} \citep{Robotham2020}, will contribute to a deep analysis. These studies will shed further light on the evolution of S0 galaxies in clusters.  

\section*{Acknowledgements}
SB acknowledges the International Macquarie University Research Training Program Scholarship (iMQRTP 2017537) for the support. M.S.O. acknowledges the funding support from the Australian Research Council through a Future Fellowship (FT140100255). JJB acknowledges support of an Australian Research Council Future Fellowship (FT180100231). JvdS acknowledges support of an Australian Research Council Discovery Early Career Research Award (project number DE200100461) funded by the Australian Government. NS acknowledges support of an Australian Research Council Discovery Early Career Research Award (project number DE190100375) funded by the Australian Government and a University of Sydney Postdoctoral Research Fellowship. The SAMI Galaxy Survey is based on observations made at the Anglo-Australian Telescope. The Sydney-AAO Multi-object Integral field spectrograph (SAMI) was developed jointly by the University of Sydney and the Australian Astronomical Observatory, and funded by ARC grants FF0776384 (Bland-Hawthorn) and LE130100198. The SAMI input catalogue is based on data taken from the Sloan Digital Sky Survey, the GAMA Survey and the VST/ATLAS Survey. The SAMI Galaxy Survey is supported by the Australian Research Council Centre of Excellence for All Sky Astrophysics in 3 Dimensions (ASTRO 3D), through project number CE170100013, the Australian Research Council Centre of Excellence for All-sky Astrophysics (CAASTRO), through project number CE110001020, and other participating institutions. The SAMI Galaxy Survey website is http://sami-survey.org/. This study uses data provided by AAO Data Central (http://datacentral.org.au/). 

%


\software{astrolibR \citep{AstroLib}, {\sc astropy} \citep{Astropy2013, Astropy2018}, Hyper-Fit \citep{Robotham2015}, K-corrections \citep{Blanton2007}, \textit{LaplacesDemon} (github.com/LaplacesDemonR), {\sc ProFit} \citep{Robotham2017}, {\sc ProFound} \citep{Robotham2018}.}



\appendix
\section{2D bulge-disk decomposition: the catalogues}
\label{2D bulge-disk decomposition: the catalogues}
The data products from the 2D photometric bulge-disk decomposition are made publicly available with this paper. The electronic formats can be found in the machine-readable version. We present one catalogue for each of the $g$-, $r$- and $i$-bands. We select the 1730 galaxies for which the pipelines {\sc ProFound}+{\sc ProFit} work in all bands. We do not include the results for the A85 galaxy sub-sample with VST/ATLAS photometric data. 

Table~\ref{rband_catalogue} presents the description of columns for the $r$-band catalogue. We list the results for the PSF, the single-component fit and the final double-component fit, specifying which photometric models are fitted. For the double-component fit we list the results for the first and second components, which are associated with the bulge and the disk. The column names related to the single/double-component results start with the corresponding band of the catalogue. In case the fits do not converge the values are replaced with -9999999. Errors equal to -1 correspond to fixed parameters. The magnitudes are not corrected for K-correction or Milky Way extinction. The last column identifies the galaxy characterization as described in Section~\ref{sec:Model selection}. Galaxies are divided into reliable double-component galaxies with $\ln$(BF)$>60$ and extra cuts applied to exclude unphysical and unreliable fits, double-component galaxies with $\ln$(BF)$>60$ and with unresolved effective radii, i.e $r_{e,disk}$ is smaller than 5\% of $r_{e,bulge}$ or $r_{e,bulge}$ is smaller than 80\% of the PSF half-width-at-half-maximum, and single-component galaxies having $\ln$(BF)$<60$ or with unphysical/unreliable double-component fits.

The description for the $g$- and $i$-bands catalogues is similar at Table~\ref{rband_catalogue}. Since our model selection is based on the $r$-band alone, the catalogues for the $g$- and $i$-bands do not report the more likely number of galaxy components. However, we report the statistical quantities of the fits such as the likelihood, reduced $\chi^{2}_{\nu}$ and Bayes Factor.

\begin{longtable}{cc}
\caption{2D bulge-disk decomposition: description of columns for the $r$-band catalogue. The full catalogues for the $g$-, $r$- and $i$-bands are available in the machine-readable versions.} 
  \label{rband_catalogue}
\\
\hline
Col. label  &  Explanation\\
\hline

cataid & Catalog ID from \citet{Owers2017}\\
photo-flag  &  Photometric data survey (SDSS; VST/ATLAS)\\
RA& Right Ascension in degrees from \citep{Owers2017}.\\
DEC& Declination in degrees from \citep{Owers2017}.\\
r-psf-model& PSF photometric model (Mof=Moffat)\\
r-psf-fwhm& PSF full-width-at-half-maximum (arcsec)\\
r-psf-con& PSF concentration\\
r-psf-axrat& PSF axial ratio (minor/major axis)\\
r-psf-ang& PSF position angle of major axis (degrees)\\
r-s-model& Single-component photometric model (Ser=S\'ersic)\\
r-s-xcen&Single-component x-position (pixels)\\
r-s-xcen-err& Error on r-s-xcen (pixels)\\
r-s-ycen&Single-component y-position (pixels)\\
r-s-ycen-err& Error on r-s-ycen (pixels)\\
r-s-mag& Single-component magnitude (mag)\\
r-s-mag-err& Error on r-s-mag (mag)\\
r-s-re& Single-component effective radius (arcsec)\\
r-s-re-err& Error on r-s-re (arcsec)\\
r-s-nser& Single-component S\'ersic index\\
r-s-nser-err& Error on r-s-nser\\
r-s-axrat& Single-component axial ratio (minor/major axis)\\
r-s-axrat-err& Error on r-s-axrat\\
r-s-ang& Single-component position angle of major axis (degrees)\\
r-s-ang-err& Error on r-s-ang (degrees)\\
r-s-ang-WCS& r-s-ang in the World Coordinate System (degrees)\\
r-s-npegged& Number of single-component parameters pegged to upper/lower fit limits\\
r-s-npegged\_warn& Number of single-component parameters pegged to upper/lower fit limits within the errors\\
r-s-redchisq& Reduced chi-square for single-component fit\\
r-s-DIC& Deviance Information Criterion for single-component fit\\
r-s-BIC& Bayesian Information Criterion for single-component fit\\
r-s-LL& Unnormalised Log-Likelihood for single-component fit\\
r-s-LML& Logarithm of the Marginal Likelihood for single-component fit\\
r-d-model1& Bulge photometric model (deV=de Vaucouleurs;simpleSer=simple S\'ersic; Ser=S\'ersic)\\
r-d\_model2&Disk photometric model (Exp=exponential)\\
r-d-xcen12& Bulge and Disk x-position (pixels)\\
r-d-xcen12-err& Error on r-d-xcen12 (pixels)\\
r-d-ycen12& Bulge and Disk y-position (pixels)\\
r-d-ycen12-err& Error on r-d-ycen12 (pixels)\\
r-d-mag1& Bulge magnitude (mag)\\
r-d-mag1-err&  Error on r-d-mag1 (mag)\\
r-d-mag2& Disk magnitude (mag)\\
r-d-mag2-err&  Error on r-d-mag2 (mag)\\
r-d-re1& Bulge effective radius (arcsec)\\
r-d-re1-err&  Error on r-d-re1 (arcsec)\\
r-d-re1-up& Upper fit limit for r-d-re1 (arcsec)\\
r-d-re2& Disk effective radius (arcsec)\\
r-d-re2-err&  Error on r-d-re2 (arcsec)\\
r-d-re2-up& Upper fit limit for r-d-re2 (arcsec)\\
r-d-nser1& Bulge S\'ersic index\\
r-d-nser1-err&  Error on r-d-nser1\\
r-d-nser2& Disk S\'ersic index\\
r-d-nser2-err&  Error on r-d-nser2\\
r-d-axrat1& Bulge axial ratio (minor/major axis)\\
r-d-axrat1-err&  Error on r-d-axrat1\\
r-d-axrat2& Disk axial ratio (minor/major axis)\\
r-d-axrat2-err&  Error on r-d-axrat2\\
r-d-ang1& Bulge position angle of major axis (degrees)\\
r-d-ang1-err&  Error on r-d-ang1 (degrees)\\
r-d-ang1-WCS& r-d-ang1 in the World Coordinate System (degrees)\\
r-d-ang2& Disk position angle of major axis (degrees)\\
r-d-ang2-err&  Error on r-d-ang2 (degrees)\\
r-d-ang2-WCS& r-d-ang2 in the World Coordinate System (degrees)\\
r-d-BT& Bulge-to-total flux ratio\\
r-d-npegged& Number of double-component parameters pegged to upper/lower fit limits\\
r-d-npegged-warn& Number of double-component parameters pegged to upper/lower fit limits within the errors\\
r-d-redchisq& Reduced chi-square for double-component fit\\
r-d-DIC &Deviance Information Criterion for double-component fit\\
r-d-BIC& Bayesian Information Criterion for double-component fit\\
r-d-LL & Unnormalised Log-Likelihood for double-component fit\\
r-d-LML& Logarithm of the Marginal Likelihood for double-component fit\\
r-DeltaBIC & Bayesian Information Criterion difference [(r-d-BIC)$-$(r-s-BIC)]\\
r-LBF & Logarithm of Bayes Factor [(r-d-LML)$-$(r-s-LML)]\\
r-ncomp & Number of galaxy components \\ 
&  [1=single; 2=double; 2b=double with (r-d-re2)$<$1.05$\times$(r-d-re1) or (r-d-re1)$<$0.9$\times$(r-psf-fwhm)]\\
\hline
\end{longtable}


\bibliography{biblio}{}
\bibliographystyle{aasjournal}



\end{document}